\documentclass[manuscript]{aastex}

\usepackage{natbib}
\slugcomment{Not to appear in Nonlearned J., 45.}

\shorttitle{Spectral Line Survey toward Molecular Cloud in the LMC}
\shortauthors{Nishimura et al.}

\begin{document}

\title{Spectral Line Survey 
       toward Molecular Clouds in the Large Magellanic Cloud}

\author{Yuri Nishimura\altaffilmark{1},
        Takashi Shimonishi\altaffilmark{2,3},
        Yoshimasa Watanabe\altaffilmark{1},
        Nami Sakai\altaffilmark{4},\\
        Yuri Aikawa\altaffilmark{5},
        Akiko Kawamura\altaffilmark{6},
        and 
        Satoshi Yamamoto\altaffilmark{1}}

\altaffiltext{1}{Department of Physics, the University of Tokyo, 
                 7-3-1, Hongo, Bunkyo-ku, Tokyo, 113-0033, Japan}
\altaffiltext{2}{Frontier Research Institute for 
                 Interdisciplinary Sciences, Tohoku University, 
                 Aramakiazaaoba 6-3, Aoba-ku, Sendai, Miyagi, 
                 980-8578, Japan}
\altaffiltext{3}{Astronomical Institute, Tohoku University, 
                 Aramakiazaaoba 6-3, Aoba-ku, Sendai, Miyagi, 
                 980-8578, Japan}
\altaffiltext{4}{RIKEN, 2-1 Hirosawa, Wako, Saitama 351-0198, Japan}
\altaffiltext{5}{Center for Computational Sciences, The University of Tsukuba, 
                 1-1-1, Tennodai, Tsukuba, Ibaraki 305-8577, Japan}
\altaffiltext{6}{National Astronomical Observatory of Japan, 
                 Osawa, Mitaka, Tokyo, 181-8588, Japan}

\begin{abstract}
Spectral line survey observations of 7 molecular clouds 
in the Large Magellanic Cloud (LMC) have been conducted 
in the 3 mm band with the Mopra 22 m telescope 
to reveal chemical compositions in low metallicity conditions.  
Spectral lines of fundamental species 
such as CS, SO, CCH, HCN, HCO$^+$, and HNC are detected in addition to those of CO and $^{13}$CO, 
while CH$_3$OH is not detected in any source and N$_2$H$^+$ is marginally detected in two sources.  
The molecular-cloud scale (10 pc scale) chemical composition is found to be similar 
among the 7 sources regardless of different star formation activities, 
and hence, it represents the chemical composition characteristic to the LMC 
without influences of star formation activities.  
In comparison with chemical compositions of Galactic sources, 
the characteristic features are (1) deficient N-bearing molecules, 
(2) abundant CCH, and (3) deficient CH$_3$OH.  
The feature (1) is due to a lower elemental abundance of nitrogen in the LMC, 
whereas the features (2) and (3) seem to originate from 
extended photodissociation regions and warmer temperature 
in cloud peripheries due to a lower abundance of dust grains 
in the low metallicity condition.  
In spite of general resemblance of chemical abundances 
among the seven sources, the CS/HCO$^+$ and SO/HCO$^+$ ratios 
are found to be slightly higher in a quiescent molecular cloud.  
An origin of this trend is discussed in relation to possible depletion 
of sulfur along molecular cloud formation.  
\end{abstract}


%%%%%%%%%%%%%%%%%%%%%%%%%%%%%%%%%%%%%%%%%%%%%%%%%%
\section{Introduction}
The Large Magellanic Cloud (LMC) is the nearest external galaxy 
($d=49.97 \pm 1.11$ kpc) \citep{pietrzynski2013eclipsing}.  
Taking advantage of proximity to the Sun, extensive studies 
have been carried out on molecular-cloud evolution and star formation 
in a galactic scale (\citet{fukui2010molecular} and references therein).  
In contrast to Galactic molecular clouds, individual distances to 
molecular clouds in the LMC are almost the same, 
which guarantees a fixed spatial resolution and thus provides 
an excellent opportunity for statistical studies of molecular clouds.  
The LMC is also interesting from an astrochemical point of view.  
The metallicity in the LMC is known to be lower than 
in the solar neighborhood by a factor 2 or more 
\citep{westerlund1997magellanic}.  
In particular, \citet{dufour1982carbon} reported that 
the C/H and N/H ratios 
are lower than those in the solar neighborhood 
by a factor of 6 and 10, respectively (Table \ref{elementalabundance}).  
These differences of the elemental abundances 
significantly should affect physical and chemical processes in molecular clouds.  
For instance, photodissociation and photoionization are expected to be 
more effective for a given H$_2$ column density 
owing to a lower abundance of dust grains in the low metallicity condition 
\citep[\emph{e.g.}][]{millar1990chemical}.  
Such effects may result in chemical compositions of molecular clouds characteristic to the LMC.  

So far, chemical compositions of molecular clouds in the LMC were 
mainly studied toward high-mass star-forming regions 
containing well-developed H II regions, such as N113 and N159.  
Bright molecular emission lines were observed in these sources, 
and abundances of various molecules derived from such observations 
are compared with those in high-mass star forming regions in our Galaxy, 
such as Orion KL and Sgr B2 
\citep{wang2009abundances, heikkila1999molecular, chin1997molecular, paron2014aste}.  
However, the above sources in the LMC involve a cluster of high-mass stars 
and have a very complex physical structure, which prevents us 
from understanding intrinsic chemical nature of molecular clouds specific to the LMC.  
For this purpose, we need to observe relatively quiescent 
molecular clouds without strong influence of cluster formation.  
Such an effort reported so far is the chemical study of the high-mass star-forming cloud, 
N44C, toward which several molecules other than CO and their isotopologues 
are detected \citep{chin1997molecular}.  
However, we still need to observe more sources in order to examine the effect of 
star-formation activities on chemistry and to extract the intrinsic chemical nature.  

Currently available single-dish telescopes located in the southern hemisphere
have a typical angular resolution of $30-40$ arcsec in the 3 mm band, 
which corresponds to a linear size of $\sim$10 pc at the distance of the LMC.  
In the single-dish observations, 
we thus look at the chemical composition averaged over the molecular-cloud scale.  
Hence, the effect of individual star-forming region on the averaged chemical composition 
is expected to be smeared out.  
This is particularly true in the 3 mm band observations, because molecular line-emission in this band 
would be dominated by extended molecular gas.  
This is in contrast to the submillimeter-wave line emission 
which preferentially traces hot and dense parts of star-forming regions.  
If so, we will be able to extract the intrinsic chemical composition specific to the LMC 
with the single-dish observation in the 3 mm band.  
Importance of such molecular-cloud scale chemical composition is now being recognized in studies 
on external galaxies \citep[\emph{e.g.}][]{watanabe2014spectral}.  
The results for the LMC will be a fundamental base for interpreting chemical compositions 
of distant galaxies observed at a high angular resolution.  

With this in mind, we conducted 3-mm spectral line survey observations 
toward 7 molecular clouds without infrared point sources in the LMC 
having different star-formation activities by using the Mopra 22 m telescope.  

\section{Source Selection \label{sourceselection}}
For the spectral line survey, we first selected candidate sources 
having different star-formation activities.  
We selected 16 sources on the basis of the CO map data (MAGMA; \citet{wong2011MAGMA}), 
which are not associated with infrared sources (`starless clouds').  
Non-detection of the point sources at 8 $\mu$m and 24 $\mu$m with \textit{Spitzer} or \textit{AKARI} 
generally means absence of high-mass young stellar objects except for very infant ones 
\citep{whitney2008spitzer, kato2012akari, gruendl2009high}.  
These `starless' sources are referred to hereafter as Category A sources in this paper.  
We also selected 17 high-mass star forming clouds which are associated 
with high-mass YSOs observed with \textit{Spitzer} or \textit{AKARI}, 
but not embedded in a prominent H II regions (referred to hereafter Category B sources) 
\citep{shimonishi2010spectroscopic, shimonishi2013akari, seale2009evolution, gruendl2009high}.  

We then observed these candidate sources in the $^{13}$CO ($J=1-0$), 
HCO$^+$ ($J=1-0$), and HCN ($J=1-0$) lines with the Mopra 22 m telescope.  
As a result, we found the bright sources in the HCO$^+$ and HCN line for each category; 
CO Peak 1 and NQC2 for Category A and N79, N44C, and N11B for Category B.  
All of the target regions are associated with relatively strong $^{12}$CO ($J=1-0$) emission. 
Based on the MAGMA CO map of the LMC, a typical CO integrated intensity of the regions 
such as N159W, N113, and N44BC, in which multiple molecular emission lines 
were detected by the SEST 15 m telescope, is $16-61$ K km s$^{-1}$.  
The typical CO integrated intensity of our target sources is $10-34$ K km s$^{-1}$, 
which is almost comparable to the above well-studied sources.  
Hence, the selected sources are likely molecular rich.  
We will publish details of the source selection and the result of the $^{13}$CO, HCO$^+$, and HCN survey 
in a separate publication \citep{shimonishiinprep}.  

In addition, we included the two active star forming regions with extended H II regions, 
N113 and N159W (referred to hereafter Category C sources), as references, in our source list.  
In total, for this study we selected 7 sources, as summarized in Table \ref{position}.  
Figure \ref{map} shows the MAGMA CO ($J=1-0$) maps and the 24 $\mu$m images 
by \textit{Spitzer} MIPS of the target sources.  
A brief description of each source is given below, 
where the cloud name is based on \citet{henize1956catalogues}, if available.

\noindent
\textbf{(Category A)}\\[-10mm]

CO Peak 1 is located at the eastern part of the LMC.  
It does not host any infrared sources, as shown in Figure \ref{map}, suggesting that 
high-mass star formation has not started yet in this cloud.  
Strong CO ($J=1-0$) emission (31.1 K km s$^{-1}$) is detected 
in the MAGMA observations.  
Deep molecular line observations are carried out for the first time in this study.  

NQC2 is a quiescent molecular cloud located at the central part of the LMC.  
This source was recognized in the \textit{Herschel} PACS map of the LMC, 
and was confirmed to have fairly strong emission of HCO$^+$ 
during the above survey \citep{shimonishiinprep}.  
It also shows a bright CO emission in the MAGMA observations, 
while it does not harbor 24 $\mu$m point sources (Figure \ref{map}).  
No molecular line observations have been reported for this source except for that of CO.  

\noindent
\textbf{(Category B)}\\[-10mm]

N79 is an H II region located in the western part of the LMC.  
In this study, we observed the relatively quiescent cloud 
not directly associated with the H II region.  
It harbors an embedded high-mass YSO which show 
dust and ice absorption bands source \citep{seale2009evolution}.  
%The spectral line survey data in 0.8 mm 
%indicates on-going high-mass star formation.  

N44C is a molecular cloud located in the central part of the LMC.  
N44C hosts an embedded high-mass YSO named ST2, 
whose luminosity is estimated to be $\sim 2\times10^5 L_{\odot}$ 
\citep{shimonishi2010spectroscopic}.
Although a few faint YSOs are involved within the observed beam, 
the total mid-infrared flux toward the N44C region is dominated 
by the single high-mass YSO, as shown in Figure \ref{map}.  
Hence, we selected this position as the Category B source.  
Detection of the dust and ice absorption bands 
in the infrared spectrum of this central YSO \citep{shimonishi2010spectroscopic} 
suggests that the N44C cloud is still in the very early stage of star formation.  
Note that molecular line observations are observed toward N44BC 
by \citet{chin1997molecular}, which is separated by $1'$ from our observed position 
and is not associated with high-mass YSOs.  
In N44BC, the lines of 8 species including HCO$^+$, HCN, HNC, CCH, CS, and SO are detected 
in the 3 mm band with the SEST 15 m telescope.  
Hence, we can compare our result toward N44C with theirs to examine the effect of star formation 
on the chemical composition.  

N11B is a member of N11 complex, the second largest H II region 
after the 30 Doradus region.  
We observed the position not directly associated with 
the extended H II region.  It involves a relatively evolved YSO, 
which shows emission lines due to PAH and ionized gas 
\citep{seale2009evolution}.  
\citet{barba2003active} also reported several embedded YSOs.  
It is also reported that the CH$_3$OH maser is associated with this YSO 
\citep{ellingsen2010masers}.  
Furthermore, a variety of far-infrared emission lines due to ionized gas 
are detected toward our beam position by \textit{Herschel} \citep{lebouteiller2012physical}.  
The above characteristics suggest that the N11B region is 
relatively evolved in comparison with the other two target sources in this category.  

\noindent
\textbf{(Category C)}\\[-10mm]

N113 is an active cluster-forming site located 
in the central port of the LMC, which aparts from the 30 Doradus region 
by about 2$^{\circ}$.  It is associated with prominent H II regions 
(NGC1874, NGC 1876, and NGC1877) \citep{bica1992bar}.  
This is a well studied source in astrochemistry; 
several molecular-line studies have been reported 
\citep[\emph{e.g.}][]{paron2014aste, wang2009abundances, chin1997molecular}.  

N159W is one of active cluster-forming regions associated with the H II region LH105 
\citep{lucke1970catalogue}.  
This source has extensively been studied as a prominent on-going 
star forming region by observations in various wavelength 
\citep[\emph{e.g.}][]{mizuno2010warm,ott2010first}.  
Molecular line observations are also reported by \citet{heikkila1999molecular, johansson1994interstellar}.  

\section{Observations}
The spectral line survey observations were carried out with the Mopra 22 m telescope of 
the Australia Telescope National Facility (ATNF) 
from June to October in 2013 and from April to October in 2014.  
We observed the 7 sources selected above, whose positions are summerized 
in Table \ref{position}.  
We used the 3 mm InP HEMT Monolithic Microwave Integrated Circuit (MMIC) receiver as a front end.  
The observed frequency range is from 85 GHz to 116 GHz.  
We observed two orthogonal polarization signals simultaneously.  
For NQC2, N79, and N113, the frequency range from 101 GHz to 108 GHz is missing.  
We did not observe this frequency range for these three sources to save the telescope time, 
because no spectral lines were detected in the other sources (Table \ref{detected}).  
The half-power beam width (HPBW) of the telescope is 
$38''$ and $30''$ at 90 and 115 GHz, respectively.  
The telescope pointing was checked by observing 
nearby SiO maser source (R Dor) every 1.5 hours, 
and the pointing accuracy was estimated to be better than $5''$.  
We used the Mopra spectrometer (MOPS) in the `wideband' mode, 
which simultaneously cover the band width of 8 GHz, as a back end.  
The frequency resolution is 270 kHz per channel, 
and we binded 3 successive channels in the analysis 
to improve the signal-to-noize ratio.  
The resultant velocity resolution is 2.73 km s$^{-1}$ at 90 GHz.
We employed the position-switching mode (OFF-ON-ON-OFF sequences) 
with individual integrations of 30 s for all the observations. 

The total on-source integration time was 16, 25, 8, 17, and 26 hours 
for CO Peak 1, NQC2, N79, N44C, and N11B respectively.  
As for the active star-forming regions, N113 and N159W, 
we conducted shorter observations.  
A typical system temperatures were ~200 K and ~600 K 
at 90 GHz and 115 GHz, respectively.  
A typical rms noize temperature in the main-beam temperature scale 
for each source and each frequency bands is summarized in Table \ref{detected}.  
It ranges from 5.9 mK to 80.9 mK, depending on the frequencies and the sources.  
The line intensity was calibrated by the chopper wheel method.  
The antenna temperature is divided by the main beam efficiency 
of 0.5 to obtain the main-beam temperature, $T_{\rm MB}$.  
The observation data were first reduced with the ATNF analysis programs, 
ASAP, and then detailed analyses were carried out using our own codes.  

\section{Results and Discussion}
\subsection{Observed Spectra and Line Parameters}
Figures \ref{spectra}a - \ref{spectra}f show compressed spectra 
from 85 GHz to 101 GHz toward 
CO Peak 1, NQC2, N79, N44C, N11B, N113, and N159W, respectively.  
In preparation of these data, we corrected slight intensity variation 
among the subscans for 1.5 hours by using the intensity of 
the strongest line of the observing band.  
For the compressed spectra, we subtracted the baseline 
by using the 5th-order polynomial over the 2.2 GHz region.  
Although the standing waves caused by the telescope system are still visible, 
we were readily able to identify emission lines, because a typical line width 
observed toward the target sources (5 km s$^{-1}$) is narrower than the period of the standing waves.  
Lines are identified with the aid of the spectral line database CDMS 
\citep{muller2001cologne, muller2005cologne}.  
We used the $^{13}$CO line velocity ($V_{\rm LSR}$) of 
229, 266, 285, 282, 232, 235 and 238 km s$^{-1}$ 
for CO Peak 1, NQC2, N11B, N44C, N79, N113, and N159W, respectively, to identify molecular lines.  
We detected 8 molecules (CO, $^{13}$CO, CS, SO, CCH, HCO$^+$, HCN, and HNC) 
in the seven sources, 
$c$-C$_3$H$_2$ in N79, N44C, N11B, N113, and N159W, 
and CN in N79, N44C, and N113 (Table \ref{lineparameters}).  
N$_2$H$^+$ is marginally detected in N113 and N159W.  
The detection criterion is that the line is detected at the expected velocity 
with the significance of 5$\sigma$ or higher in the integrated intensity.  
Note that the $c$-C$_3$H$_2$ line was not covered 
with the frequency setting for CO Peak 1.  
Since line intensities in CO Peak 1 and NQC2 are 
typically weaker by a factor of 2 than the other sources, 
a fewer lines were detected in these sources.  
The signal-to-noise ratio is rather poor for N113 and N159W 
in comparison with the previous studies 
\citep{wang2009abundances, heikkila1999molecular, chin1997molecular}, 
and hence, only strong lines were detected in this study.  

The spectral profile of each molecular line is 
shown in Figures \ref{eachline1} and \ref{eachline2}.  
For Figures \ref{eachline1} and \ref{eachline2} and Table \ref{lineparameters}, 
the baselines are subtracted in a narrow range (typically 100 MHz) 
by using the 5th-order polynomials to remove the base line ripples.  
We conducted Gaussian fitting to each base-line-subtracted spectral line 
to obtain the $V_{\rm LSR}$, $\Delta v$, and the peak $T_{\rm MB}$, 
where the observed frequencies are converted to the LSR velocities 
based on the rest frequency of each transition given in CDMS.  
Derived line parameters are summarized in Table \ref{lineparameters}. 

Although the seven sources have different star-formation activities, 
as mentioned in Section \ref{sourceselection}, 
the spectral patterns of the seven sources look similar to one another, 
except for a slight difference discussed later.
In order to reveal the similarity, we inspected the molecule-to-molecule correlation 
of the integrated intensities of the 7 sources.  
The correlation coefficient $c$ is calculated as 
$$
c = \frac{\sum(x_i-\bar{x})(y_i-\bar{y})}{\sqrt{\sum(x_i-\bar{x})^2\sum(y_i-\bar{y})^2}} 
$$
where $x_i$ and $y_i$ are observed integrated intensities for the $i$-th source.  
$\bar{x}$ and $\bar{y}$ are the averages of $x_i$ and $y_i$, respectively.  
If the correlation is good, the relative abundance between the two species 
is similar despite the different column densities along the line of sight.  
Table \ref{correlationcoefficient} shows the results for the 7 species.  
The correlation coefficients are generally higher than 0.9, except for the $^{13}$CO, 
indicating that the relative abundances of any pair of CCH, HCN, HCO$^+$, HNC, CS, and SO 
are similar among the 7 sources.  
On the other hand, the correlation coefficients between $^{13}$CO and the other species are lower than 0.8.  
Thus the abundances of CCH, HCN, HCO$^+$, HNC, CS, and SO relative to $^{13}$CO show more 
source-to-source variation.  
This implies that the emmiting region of the $^{13}$CO line 
is different from those of the other species, 
because of its lower critical density and/or saturation effect.  

In this study, we observed the position involving the bright 24 $\mu$m source for N44C.  
On the other hand, \citet{chin1997molecular} observed various molecular lines 
toward the position offset by $1'$ with the SEST 15 m telescope, 
whose beam does not involve the bright 24 $\mu$m source.  
We confirmed that the spectral pattern constructed from the line intensity data observed 
by \citep{chin1997molecular} is very similar to our result.  
This fact further suggests that star-formation activities would not seriously affect 
the chemical compositon at a 10 pc-scale.  

Thus, an important result of this study is that the observed spectral pattern 
with Mopra at 3 mm is similar for all the observed sources, 
although these sources have different star formation activities.  
The previous studies focus on particular sources (\emph{e.g.} N113 and N159W) 
with higher sensitivity and often with higher excitation lines.  
As a result, the chemical composition derived from these studies 
is more or less specific to each source.  
In contrast, we are extracting a common feature of the chemical composition 
averaged over the molecular cloud scale (10 pc scale) with Mopra.  
Such a chemical composition is almost free from the influence of individual star formation activities, 
and can be used to study the characteristic chemical composition specific to low-metallicity conditions.  

A similar result for molecular-cloud scale 
chemical compositions in a much larger scale (kpc scale)
is recently reported for the spiral arm clouds of 
the external galaxy M51 \citep{watanabe2014spectral}.  
They observed the two positions, P1 and P2, in the spiral arm of M51, 
which are separated by $20''$, with the IRAM 30 m telescope.  
Although the star formation efficiency evaluated from the H$\alpha$ and 24 $\mu$m emission is 
higher in P1 than in P2 by a factor of 1.5, 
the chemical composition is essentially the same in the two positions.  
Thus, local star formation and its feedback do not strongly affect 
the spectral pattern observed at a resolution of a molecular-cloud scale.  

\subsection{Comparison with Galactic Molecular Cloud}
Comparison of our result in the LMC with the Galactic sources is not easy.  
Apparently, the observed spectral pattern does not originate from star forming cores, 
but reflects the average over the molecular cloud scale.  
In fact, the spectral pattern observed in this study is much different 
from the Galactic star forming regions 
such as Orion KL and NGC 2264 (Watanabe et al. 2015) (Figure \ref{spectra}).  
Hence, we need the chemical composition of the Galactic sources 
averaged over the molecular cloud scale (10 pc) for fair comparison.   
However, such data are not available, except for the Galactic center region 
which has peculiar physical and chemical conditions 
\citep[\emph{e.g.}][]{jones2012spectral, jones2013spectral}.  
Although molecular clouds harbors dense cores and star forming cores, 
most parts of them would be less dense.  
Hence, it is worth comparing our result in the LMC with the chemical composition 
of the translucent clouds as the best possible effort.  

Turner and his collaborators observed the 
HCO$^+$, HCN, HNC, CS and SO lines toward many translucent clouds 
with the total A$_{\rm v}$ of about $2-4$ magnitude and 
H$_2$ density of $10^3-10^4$ cm$^{-3}$
\citep{turner1994physics, turner1995physicsA, turner1995physicsB, turner1996physics, turner1997physics, turner1998physics, 
turner1999physics, turner2000physics}.
We averaged the intensities of 38 translucent clouds, 
and prepared a hypothetical spectrum of a translucent cloud, 
as shown in Figure \ref{spectra}i.  

The hypothetical spectrum is compared with the observed spectra in the LMC.  
The intensities of HCN and HNC relative to that of HCO$^+$ 
are apparently brighter in the hypothetical cloud than the spectra in the LMC clouds.  
This is due to the lower elemental abundance of nitrogen in the LMC 
than in the Galaxy (Table \ref{elementalabundance}), as described in section \ref{abu}.  
Except for the difference of N-bearing species, 
the LMC spectra seem to resemble the hypothetical spectrum of 
the translucent cloud.
Note that molecules other than HCN, HCO$^+$, HNC, CS, and SO lines 
are not included in the hypothetical spectrum 
due to the lack of the available data for the translucent clouds.  

In order to see whether such comparison with the hypothetical translucent cloud is meaningful, 
we compare the hypothetical translucent cloud spectrum 
with the spectrum taken toward the spiral arm of M51 (Watanabe et al, 2014), 
which has a similar metallicity to the solar neighborhood.  
Since the H$_2$ density derived from the two transitions of H$_2$CO 
($1_{01}-0_{00}$ and $2_{02}-1_{01}$) by the statistical equilibrium calculation 
under the large-velocity-gradient (LVG) approximation \citep{goldreich1974molecular} 
is $\sim10^4$ cm$^{-3}$ in the spiral arm of M51 \citep{nishimurainprep}, 
the contribution of less dense parts of molecular clouds is dominant.  
The hypothetical spectral pattern of the Galactic translucent cloud looks similar to that observed in M51.  
Hence, it is most likely that the lower HCN and HNC abundances relative to HCO$^+$ in the LMC clouds 
originate from the low elemental abundance of nitrogen.  

\subsection{Molecular Abundances \label{abu}}
In this section, we evaluate beam-averaged column densities of HCN, HCO$^+$, HNC, CS, and SO.  
Considering that we are looking molecular cloud scale (10 pc) chemical composition 
in our observations of the LMC, we assume the H$_2$ density to be $5\times10^3-5\times10^4$ cm$^{-3}$.  
In this condition, most of the observed lines except for $^{13}$CO are sub-thermally excited, 
and the level population would naturally be in the non-LTE condition.  
Hence, it is safe to use the LVG statistical equilibrium calculation to derive the column densities
\citep{goldreich1974molecular}.  
Drawback is that we have to assume the density and the temperature.  
However, these two parameters can be constrained within reasonable ranges.  
The H$_2$ density is reported to be of the order of $10^4-10^6$ cm$^{-3}$ for N113 and N159W
\citep{wang2009abundances, heikkila1999molecular}.  
Since these are mostly derived by using the higher excitation lines with high critical density, 
the emitting regions of the 3 mm lines observed in this study would originate from less dense regions.  
Hence, we employed the H$_2$ density range mentioned above.  
The temperature range is set from 10 K to 30 K, according to the result of \citet{heikkila1999molecular}.  
Under these assumptions, we calculated the column density of each molecule 
directly from the observed intensity.  
We basically employed the RADEX code \citep{schoeier2005atomic} 
for this purpose except for CCH: we prepared the statistical equilibrium calculation code 
for CCH by employing the collisional cross sections reported in RADEX.  
The results are summarized in Table \ref{columndensity}.  
Although the column densities are sensitive to the assumed H$_2$ density and temperature, 
the abundance ratios such as HCN/HCO$^+$ are much less sensitive.  
The abundances relative to HCO$^+$ are shown in Table \ref{moleculeratio}.  
Note that these results does not vary significantly, 
if we assume the H$_2$ density of 10$^5$ cm$^{-3}$.  
We evaluated the error from the range of deviation due to different physical conditions 
as well as the rms noise of the spectra.  
For N113 and N159W, the derived abundance relative to HCO$^+$ are almost consistent with the values 
calculated from the previous reports, as shown in Table \ref{moleculeratio}.  

\subsubsection{Nitrogen Bearing Molecules}
One of the characteristic features in the chemical compositions 
in the LMC is the deficiency of the N-bearing molecules due to 
the low elemental abundance of nitrogen, which is already evident in the spectral pattern.  
We first discuss this feature by focusing on the HCN/HCO$^+$ and HNC/HCO$^+$ ratios.  
The HCN/HCO$^+$ and HNC/HCO$^+$ ratio averaged for the 7 sources are $3.4\pm0.6$ and $0.8\pm0.2$, 
respectively.  For quantitative comparison, 
we derived the column densities of HCO$^+$, HCN, and HNC for the hypothetical translucent cloud, 
and evaluated the HCN/HCO$^+$ and HNC/HCO$^+$ ratios 
to be 9.3 and 2.2 (Table \ref{moleculeratio}), respectively. 
If the brightest cloud (CB17) among the Galactic translucent clouds 
in the sample mentioned above is dropped, 
the ratios are 19.3 and 3.5, respectively.  
The elemental N/O ratio in the LMC is lower by a factor of 3.4 
than that in the solar neighborhood (Table \ref{elementalabundance}).  
The HCN/HCO$^+$ and HNC/HCO$^+$ ratios in the LMC are also lower 
by a similar factor than in the Galactic translucent clouds (Table \ref{moleculeratio}).  
Hence, it is most likely that the deficiency of the N-bearing molecules 
originates from the elemental deficiency of nitrogen in the LMC.  

A similar comparison is also carried out for M51 P1.  
We evaluated the HCN/HCO$^+$ and HNC/HCO$^+$ ratios in M51 P1 
by using the statistical equilibrium calculations to be 9.6 and 1.7, respectively, 
for a consistent comparison with the analyses of the LMC (Table \ref{moleculeratio}).  
These ratios are larger than those in the LMC.  
This result further suggests that the HCN/HCO$^+$ and HNC/HCO$^+$ ratios are 
related to the elemental abundance of nitrogen.  

\citet{millar1990chemical} carried out 
chemical model calculations of dark clouds in the LMC, 
comparing with chemical compositions of our Galactic molecular clouds.  
They reported that the abundance of N-bearing species 
are sensitive to the elemental abundance of N.  
Specifically, they took account of the lower grain abundance 
and the higher UV radiation in the LMC than in our Galaxy, 
and conducted simulations for two types of the elemental abundances; 
the N/O ratios are 0.3 times (model L1) and 0.2 times (model L2) than those in our Galaxy.  
The resultant abundance ratios are 2.3 and 0.1 for HCN/HCO$^+$ 
and 1.3 and 0.09 for HNC/HCO$^+$ in their model L1 and L2, respectively, at the early time.  
These are lower than the HCN/HCO$^+$ and HNC/HCO$^+$ ratios calculated for the Galactic conditions, 
5.7 and 3.1, respectively.  
The low abundances of HCN and HNC are thus related to the low elemental abundance of nitrogen.  

Next, we focus on the HNC/HCN ratio.  
HNC is a geometrical isomer of HCN, and is less stable by 0.49 eV \citep{bentley1993highly}.  
The HNC/HCN ratio reflects production and isomerization mechanisms 
of these two species in molecular clouds \citep[\emph{e.g.}][]{hirota1998abundances}.  
We evaluated the HNC/HCN ratio in the LMC to be $0.26\pm0.08$, which is lower 
than the ratio in typical dark clouds ($0.54\sim4.5$, \citet{hirota1998abundances}).  
It is close to the ratio reported for some Galactic diffuse clouds, where the HCN and HNC lines 
are detected against the continuum sources ($0.21\pm0.05$, \citet{liszt2001comparative}).  
\citet{hirota1998abundances} reported that the HNC/HCN ratio 
is lower under higher-temperature environments: the ratio decreases above 24 K.  
The relatively low ratios observed in the LMC clouds 
may originate from warmer temperature conditions than in the Galactic 
translucent clouds due to higher UV field and/or lower grain abundance in the LMC.  
\citet{wang2009abundances} pointed out that the lower ratio of HNC/HCN 
in N113 reflects the effect of strong UV radiation, on the basis of 
the observation of 8 $\mu$m emmision of PAH \citep{wong2006synthesis}.  
In this study, we found the HNC/HCN ratio is lower 
even in the quiescent clouds (Category A and B).  
Our result suggests that the warmer temperature caused by the lower abundance of dust grains 
in the low metallicity conditions would be responsible for the lower HNC/HCN ratio (Section \ref{CH3OH}).  

Finally, we briefly discuss non-detection of N$_2$H$^+$ 
except for N113 and N159W in this study.  
This seems to originate mainly from the low elemental abundance of nitrogen.  
In addition, the warm condition would contribute to diminishing the N$_2$H$^+$ abundance.  
N$_2$H$^+$ is easily destroyed by the proton-transfer reaction with CO
\citep{bergin2002depletion, jorgensen2004imaging}.  
Since CO is not well depleted in warm peripheries ($T>20$ K), 
the N$_2$H$^+$ abundance would further decrease by this mechanism.  

\subsubsection{CCH}
Another characteristic feature is a relatively high abundance of CCH 
in comparison with the Galactic translucent clouds.  
We calculated the abundance ratio of CCH/HCO$^+$ for the seven observed clouds in LMC, 
and also for three translucent cloud (CB17, CB24, CB228) observed by Turner, 
under the assumption of the H$_2$ density of $5\times10^3 - 5\times10^4$ cm$^{-3}$ 
and the gas kinetic temperature of $10-30$ K, as shown in Table \ref{moleculeratio}.  
The ratio of CCH/HCO$^+$ is higher in the LMC ($\sim$12) 
than in the Galactic translucent clouds ($\sim$4.2) by a factor of 3.  
Note the elemental C/O ratio is lower in the LMC ($\sim$0.33) 
than in the solar neighborhood of our Galaxy ($\sim$0.60) by a factor of 0.55.  
If this difference of the C/O ratio is taken into account, 
the CCH/HCO$^+$ ratio in the LMC is higher than the ratio expected 
from the elemental C/O ratio by factor of 5.  

It is generally thought that CCH is abundant in the photodissociation region (PDR) 
illuminated by UV radiation \citep[$e.g.$][]{martin2014chemistry}.  
Since the metallicity in the LMC is lower than our Galaxy, 
the abundance of dust grains is accordingly lower.  
Hence, the extinction of the UV radiation by dust grains is less effective 
for a given column density of H$_2$.  
For this reason, the PDR tends to be extended deeper into the clouds 
in the LMC, which would be responsible to the relatively 
high abundance of CCH.  
In the PDR, growth of large carbon-chain molecules are suppressed 
by competitive photodissociation process.  
In fact, the abundance ratio of $c$-C$_3$H$_2$ relative to CCH is lower 
in the LMC than in the solar neighborhood of our Galaxy; 
we calculated the abundance ratio of CCH/$c$-C$_3$H$_2$ to be $\sim$10 
and $\sim$3 in the LMC and the Galactic translucent clouds, respectively.  

\subsubsection{CH$_3$OH \label{CH3OH}}
It is worth noting non-detection of CH$_3$OH in our survey.  
We obtained an upper limit of CH$_3$OH intensity and column density 
in the seven clouds in the LMC.  
The abundance ratio of CH$_3$OH/HCO$^+$ in the LMC are comparable to 
or lower than in the Galactic translucent clouds.  
This result can also be interpreted in terms of a stronger UV effect.  
CH$_3$OH is thought to be produced by hydrogenation of CO 
on dust grains, and is liberated into the gas phase by thermal 
and/or non-thermal desorption \citep[\emph{e.g.}][]{watanabe2002efficient}.
Since the adsorption temperature of CO is 20 K, 
CO is not well depleted onto dust grains in the warm condition, 
where the UV heating is effective.  
Furthermore, an evaporation time of the H atom rapidly becomes shorter 
for higher temperature, giving a less chance to react with CO.  
Hence, CH$_3$OH is not formed efficiently in warm conditions as those present in the LMC.  
This is also suggested by \citet{shimonishi2015submitted} 
on the basis of the infrared observation of solid CO$_2$.  
The dust temperature of the LMC is generally higher than that of our Galay, 
according to the infrared observations \citep{aguirre2003spectrum, sakon2006properties}, 
which is supposed to be caused by the strong UV radiation field \citep{bel1986chemical}.  
However, the present upper limits of CH$_3$OH/HCO$^+$ are not stringent enough 
to conclude this effect.  
Further sensitive observations are necessary to confirm 
the deficiency of CH$_3$OH in the LMC.  

\subsection{Abundant S-bearing Molecules in CO Peak 1 \label{S-bearing}}
Although the spectra observed toward the seven sources in the LMC 
are similar to one another as a whole, 
we find a slight but significant difference of the CS and SO abundance in CO Peak 1.  
Figure \ref{correlation} shows correlation diagrams of 
integrated intensities for the five species (HCO$^+$, HCN, HNC, CS, and SO) 
for the three sources (N113, N44C, and CO Peak 1).  
The sulfur-bearing species (CS and SO) tend to be abundant in CO Peak 1.  
To confirm this trend, we evaluated the CS/HCO$^+$ and SO/HCO$^+$ 
abundance ratios for the seven LMC clouds from the column densities.  
The results are shown in Figure \ref{s-species1}.  
The abundances of CS and SO relative to HCO$^+$ tend to be higher in CO Peak 1 
by factor of 2 than the averages for the other 6 sources.  
Star-formation activities would enhance the abundance of 
the sulfur-bearing molecules, because it is generally thought that 
liberation of sulfur atoms and/or sulfur-bearing molecules by outflow shocks 
and photostellar heating is responsible for the enhancement 
\citep[$e.g.$][]{wakelam2005sulphur, wakelam2011sulfur}.  
However, the slight enhancement of CS and SO relative 
to HCO$^+$ in CO Peak 1 is not an effect of star-formation activity, 
because CO Peak 1 does not harbor high-mass YSOs.  
In our observations of the seven sources, the abundances of sulfur-bearing species 
are the highest in the quiescent cloud, CO Peak 1.  
We here consider the two possible reasons for this result.  

First, the trend may be caused by depletion of S in evolved molecular clouds, 
or by deficiency of HCO$^+$ in young molecular clouds.  
In molecular clouds, sulfur is believed to be heavily depleted on dust grains.  
In chemical models, the gas-phase abundance of sulfur is usually 
assumed to be only 1/100 of the total elemental abundance 
\citep[$e.g.$][]{wakelam2011sulfur, aikawa2008molecular, 
leung1984synthesis, graedel1982kinetic}.  
On the other hand, it is observationally known that sulfur is much less depleted 
in diffuse clouds than in molecular clouds 
\citep{lehner2004c, van1988galactic, morton1974interstellar}.  
Hence, sulfur would be depleted onto dust grains 
along molecular-cloud formation.  
The higher CS/HCO$^+$ and SO/HCO$^+$ ratio in CO Peak 1 
than in more evolved clouds suggests that sulfur depletion is not yet completed in CO Peak 1.  
Because of the low metallicity, extinction of UV and visible light 
is less effective in the LMC than in our Galaxy for a given H$_2$ column density, 
and the dust temperature of the molecular clouds (in particular the temperature of their peripheries) 
is expected to be higher in the LMC than in our Galaxy 
\citep{aguirre2003spectrum, sakon2006properties}.  
The adsorption temperature of sulfur is estimated to be about 20 K 
by using the adsorption energy of 1100 K \citep{ruffle2000models}.  
Hence, the sulfur depletion would be less efficient for molecular clouds in the LMC.  
As a result, sulfur-bearing molecules would still be abundant in the gas phase.  
The decrease of the CS/HCO$^+$ and SO/HCO$^+$ ratios may reflect 
gradual depletion of sulfur along the evolution of molecular clouds.
However, it should be noted that such enhancement of sulfur-bearing molecules 
is not apparently seen in another quiescent cloud, NQC2.  
Hence, this trend may be specific to CO Peak 1, or NQC2 may be more evolved than CO Peak 1.  

Secondly, the above results is due to the low abundance of HCO$^+$ in CO Peak 1.  
If the average density is relatively lower in CO Peak 1 
than the other sources, the ionization degree is expected to be higher 
in CO Peak 1 due to contribution of photoionization.  
Since HCO$^+$ is produced by the reaction of ${\rm H_3}^+ + {\rm CO}$, 
its formation rate depends on the ${\rm H_3}^+$ ion abundance.  
The enhancement of ionization degree (\emph{i.e.} the electron abundance) 
by the photoionization leads to a lower abundance of ${\rm H_3}^+$, 
and thereby a lower abundance of HCO$^+$.  
Above all, the HCO$^+$ abundance may be lower in CO Peak 1, 
which would also contributes to the trend observed 
in Figure \ref{s-species1}.  

Relative contributions of the above two possibilities are difficult 
to be discriminated only from the present data.  
It is worth noting that the CS/HCN ratio seems to show 
a similar trend to the CS/HCO$^+$ ratio in spite of large uncertainties 
(Figure \ref{s-species2}).  
This suggests that the first possibility may work at least partly.  
Although further statistical studies are necessary 
for a definitive conclusion, the CS/HCO$^+$ and SO/HCO$^+$ ratios may be used 
as an evolutionary tracer of molecular clouds.

\section{Summary}
In this paper, we have revealed the chemical composition averaged over 
the molecular-cloud scale on the basis of the spectral line surveys 
in the 3-mm band toward 7 molecular clouds in the LMC.  
The results are summarized as follows: 

\begin{enumerate}
\item Eight molecular species, CCH, HCN, HCO$^+$, HNC, 
CS, SO, $^{13}$CO, and $^{12}$CO were identified in the 7 sources.  
Some other species ($c$-C$_3$H$_2$, N$_2$H$^+$, CN) were also detected in several sources.  

\item Although CO Peak 1, NQC2, N79, N44C, N11B, N113, and N159W have different 
star formation activities, their spectral patterns are found to resemble one another.  
The chemical compositions averaged over a whole moleuclar-cloud scale ($\sim10$ pc scale) 
are scarcely affected by local star-formation activities. 

\item In comparison with the hypothetical spectrum of 
a translucent cloud in our Galaxy prepared from the literature data, 
the intensities of HCN and HNC, relative to the HCO$^+$, 
are much weaker in the three LMC cloud.  
This seems to originate from the low elemental abundance of nitrogen in the LMC.  

\item In the 7 sources of the LMC, 
CCH is found to be abundant and CH$_3$OH is deficient in comparison with our Galaxy.  
These are interpreted as effects of the strong UV radiation, 
caused by the lower abundance of dust grains in the low metallicity condition.  

\item We find a slight but significant difference 
in abundance of the sulfur-bearing species relative to HCO$^+$ 
among the seven sources.  
The abundance ratios of sulfur-bearing species are highest 
in the starless cloud (CO Peak 1).  
The decrease of the ratio toward star-forming molecular clouds may reflect 
the evolution of molecular clouds.  

\end{enumerate}

\acknowledgments
We are grateful to the anonymous reviewer for valuable comments and suggestions.  
We thank the staff of the Mopra telescope for excellent support.  
This study is supported by Grants-in-Aid from Ministry of Education, Sports, 
Science, and Technologies of Japan (21224002, 25400223, and 25108005).  
YN is supported by Grant-in-Aid for JSPS Fellows (268280).

\clearpage

%%%%%%%%%%%%%%%%%%%%%%%%%%%%%%%%%%%%%%%%%%%%%%%%%%
\begin{figure}
\includegraphics[width=\hsize, bb= 0 0 530 516]{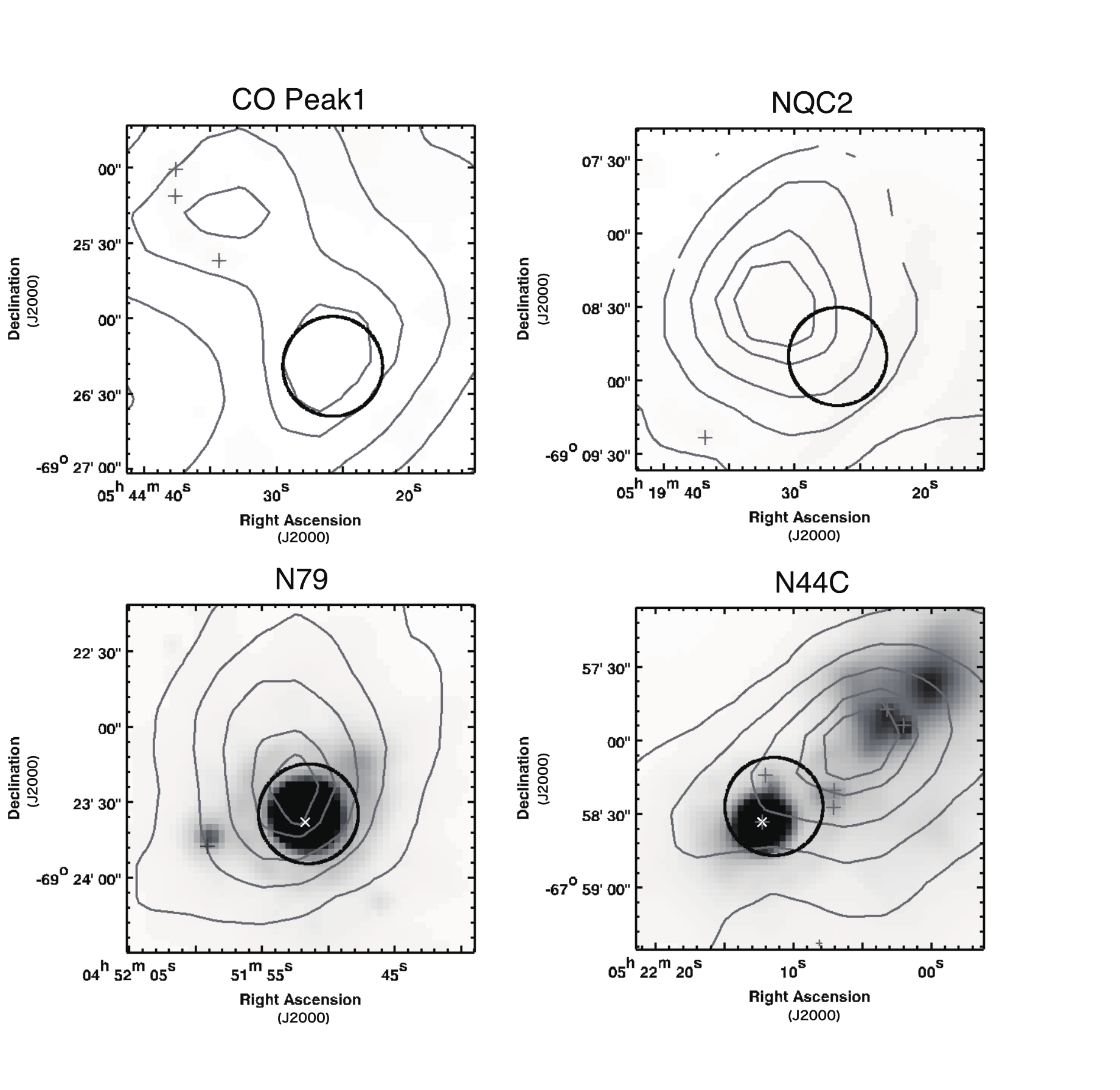}
\caption{CO ($J=1-0$) contour (MAGMA) 
and 24 $\mu$m image (\textit{Spitzer} MIPS) of 
our target molecular clouds.  
The solid circle represents the beam size of the Mopra telescope in 3 mm 
($38''\sim 10$ pc at the LMC) centered at the observed position.  
The grey crosses indicate the positions of YSOs candidates reported in 
\citet{gruendl2009high}, 
while the white x-marks indicate the positions of 
particularly bright high-mass YSOs within or near the beam position.  
\label{map}}
\end{figure}

\setcounter{figure}{0}

\begin{figure}
\includegraphics[width=\hsize, bb= 0 0 531 516]{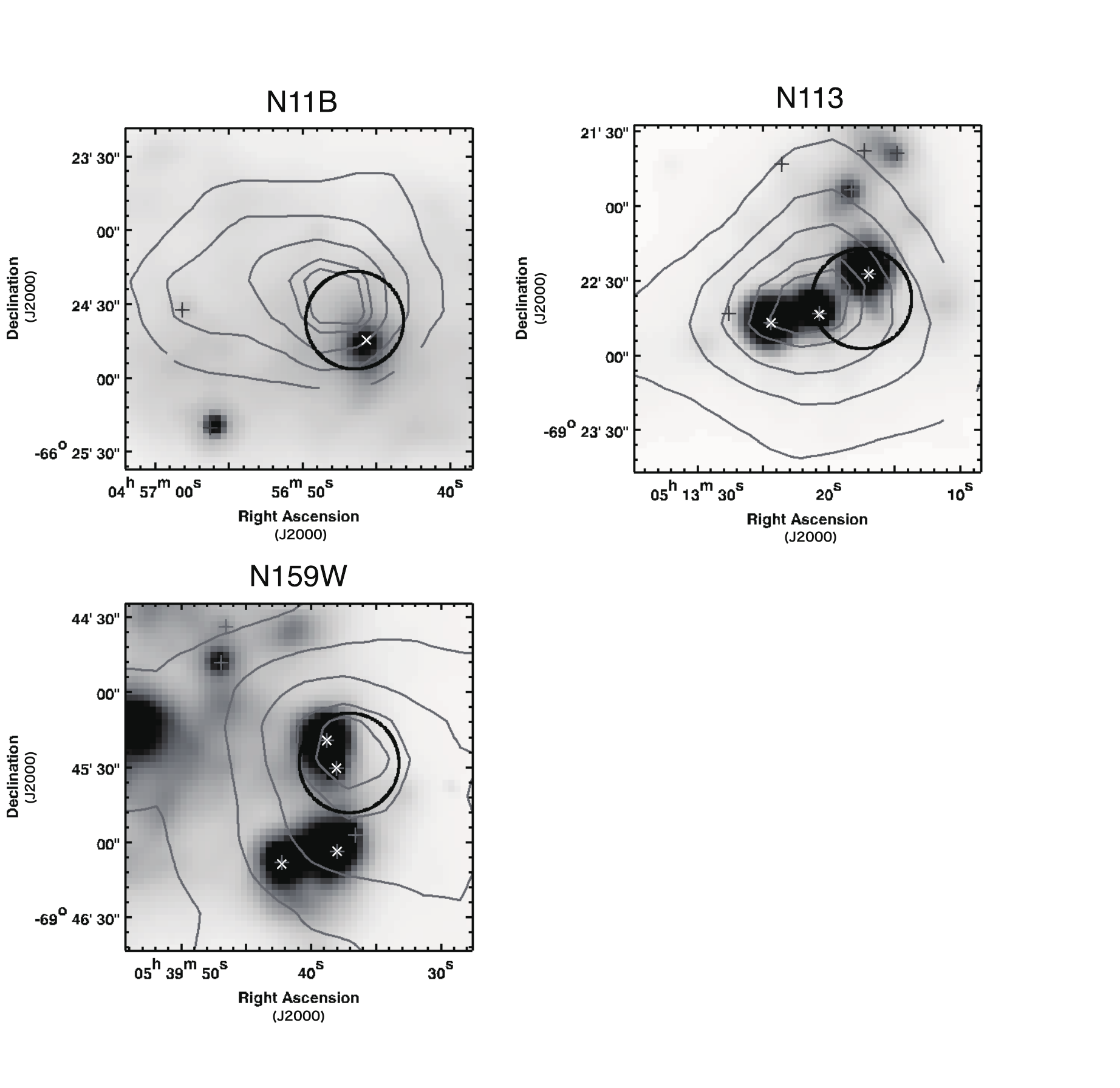}
\caption{Continued}
\end{figure}

\begin{figure}
\includegraphics[width=0.9\hsize, bb= 0 0 540 710]{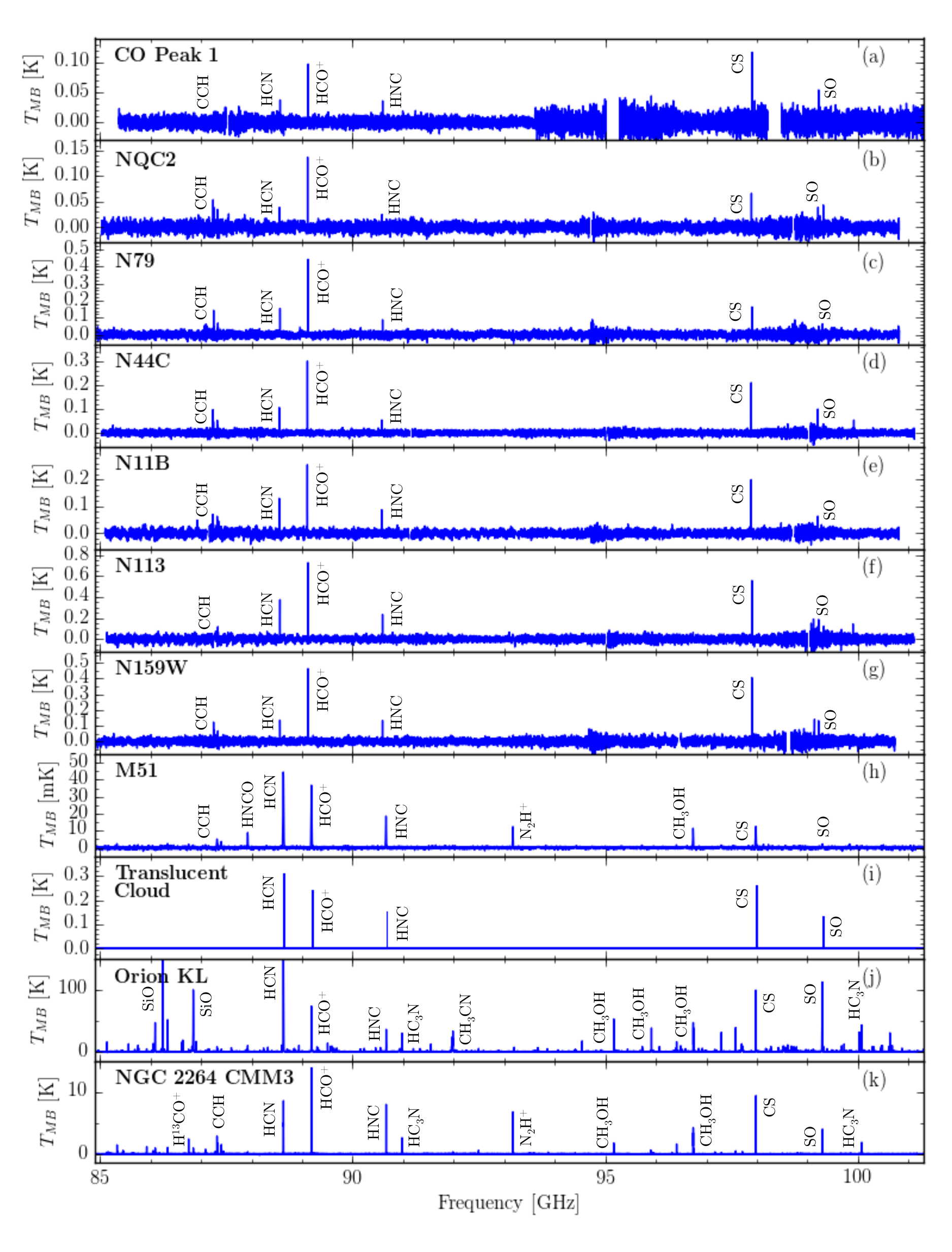}
\caption{Compressed spectra observed toward CO Peak 1 (a), 
NQC2 (b), N79 (c), N44C (d), N11B (e), N113 (f), N159W (g).  
The spectrum of the spiral arm of M51 is taken from \citet{watanabe2014spectral} (h).
The composite spectrum of a translucent cloud is prepared by averaging the 
intensities of HCN, HCO$^+$, HNC, CS and SO observed toward 38 translucent clouds 
by \citet{turner1995physicsA, turner1995physicsB, turner1996physics, 
turner1997physics} (i).  
The spectrum of Orion KL (j) and NGC 2264 CMM3 (k) 
are taken from \citet{watanabe2015inpress}.
Note that the vertical scale is different from source to source.  
\label{spectra}}
\end{figure}

\begin{figure}
\includegraphics[width=1.1\hsize, bb= 0 0 1080 576]{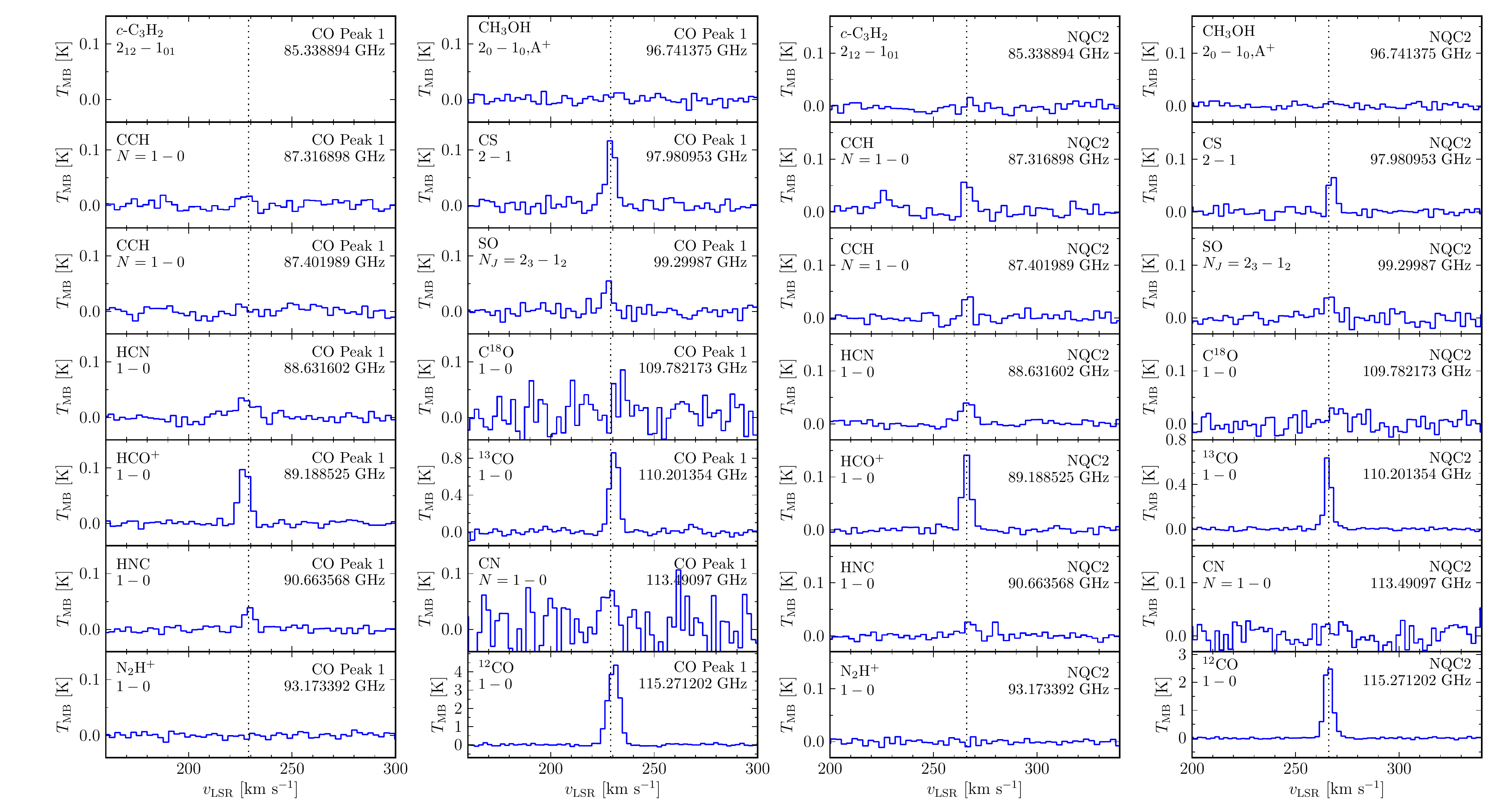}\\
\includegraphics[width=1.1\hsize, bb= 0 0 1080 576]{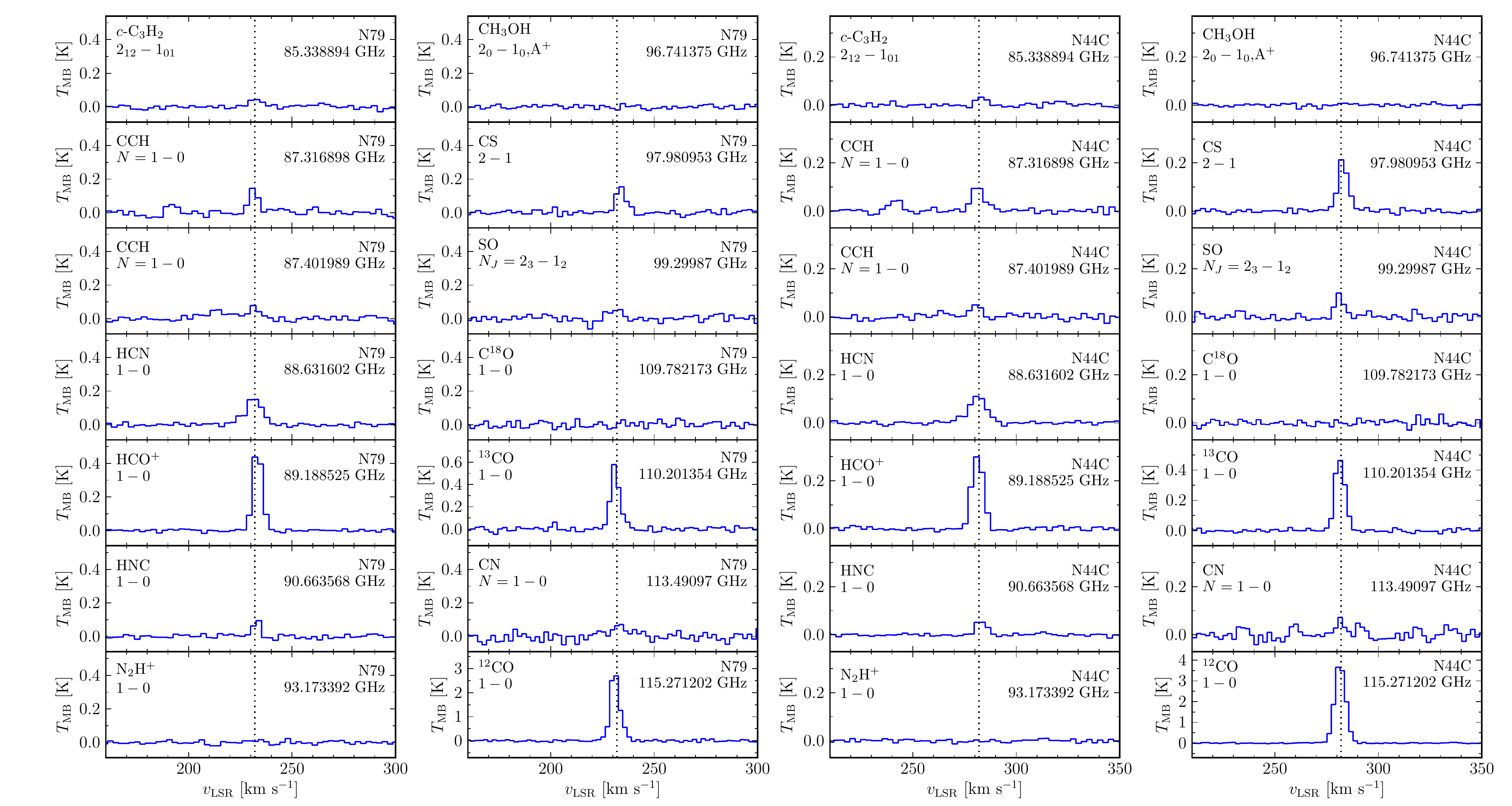}
\caption{Spectral line profiles of individual molecular transitions observed in CO Peak 1, 
NQC2, N79, and N44C.  \label{eachline1}}
\end{figure}
\begin{figure}
\includegraphics[width=1.1\hsize, bb= 0 0 1080 576]{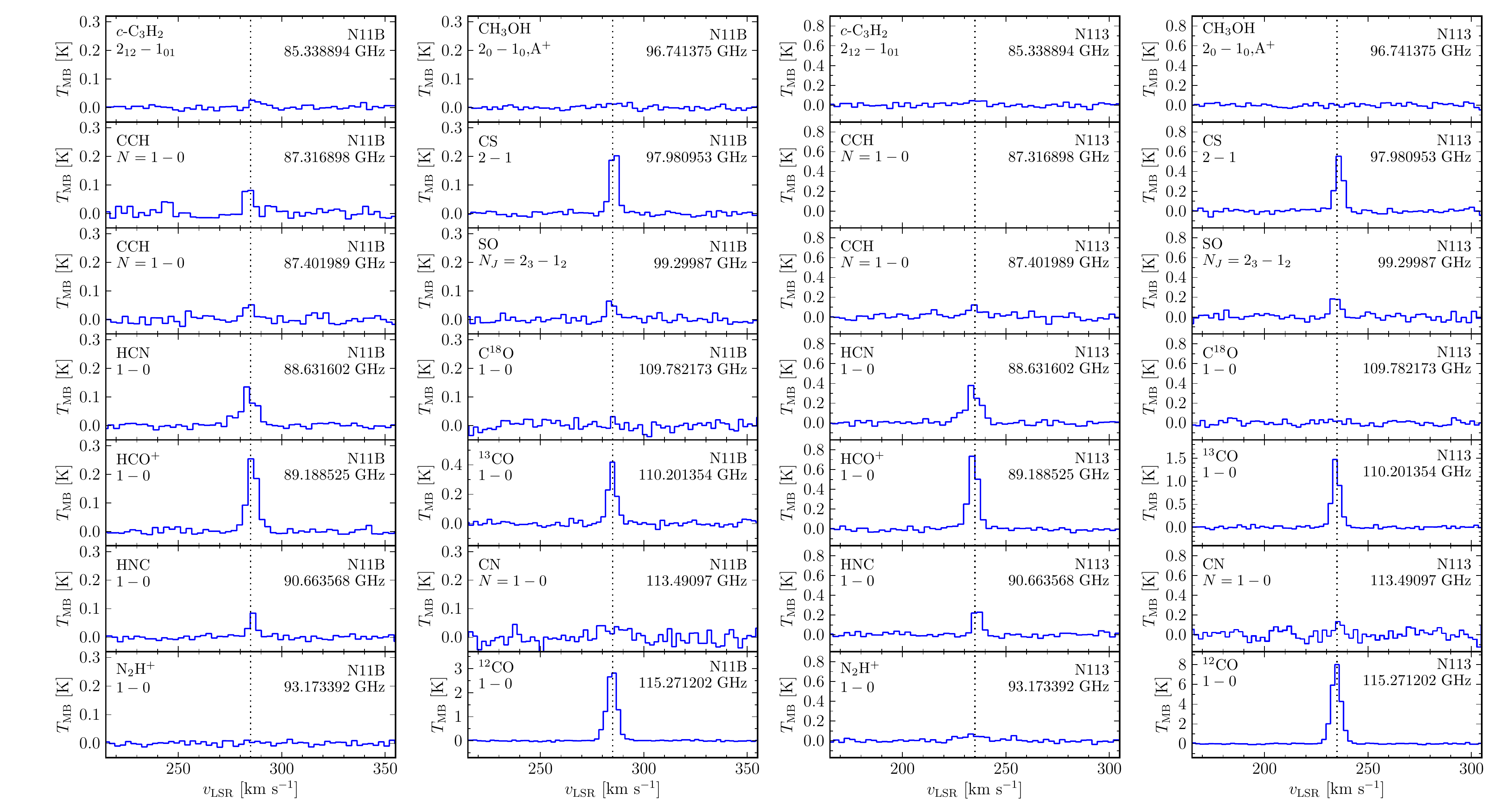}\\
\includegraphics[width=1.1\hsize, bb= 0 0 1080 576]{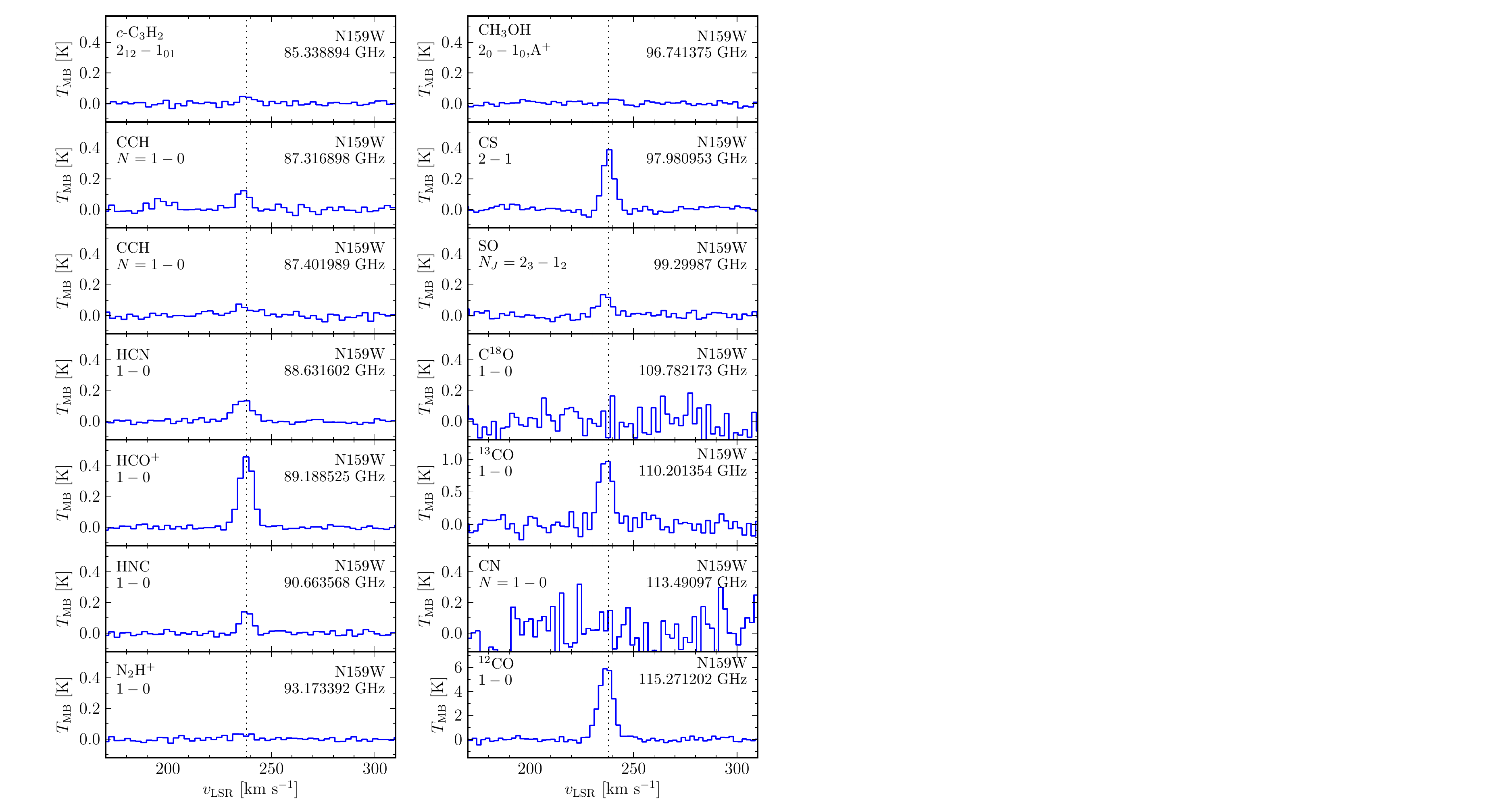}
\caption{Spectral line profiles of individual molecular transitions observed in N11B, 
N113, and N159W.  \label{eachline2}}
\end{figure}

\begin{figure}
\includegraphics[width=\hsize, bb= 0 0 936 339]{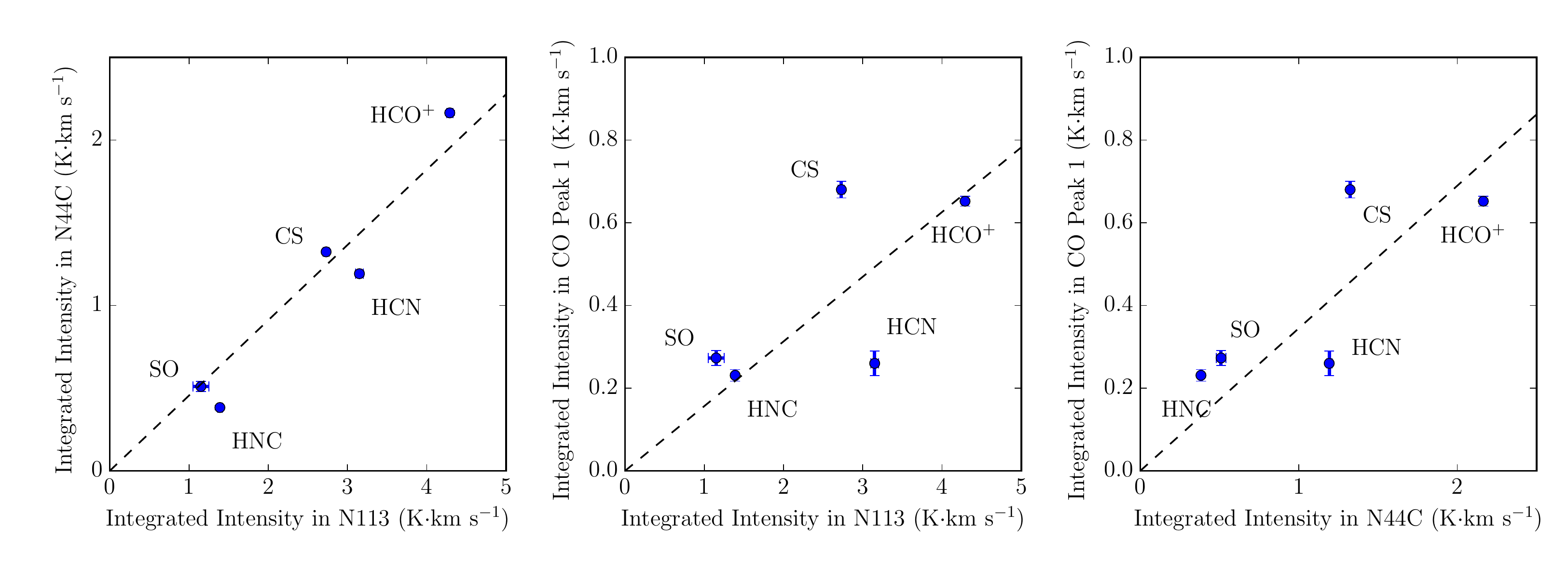}
\caption{Correlation diagrams of integrated intensities of detected species 
between N113 and N44C (left), N113 and CO Peak 1 (middle), N44C and CO Peak 1(right).  
The dashed line indicates the average ratio of the integrated intensities of the two sources.   
The CS and SO lines tend to be relatively stronger in CO Peak 1.  \label{correlation}}
\end{figure}

\begin{figure}
\includegraphics[width=0.6\hsize, bb= 0 0 576 410]{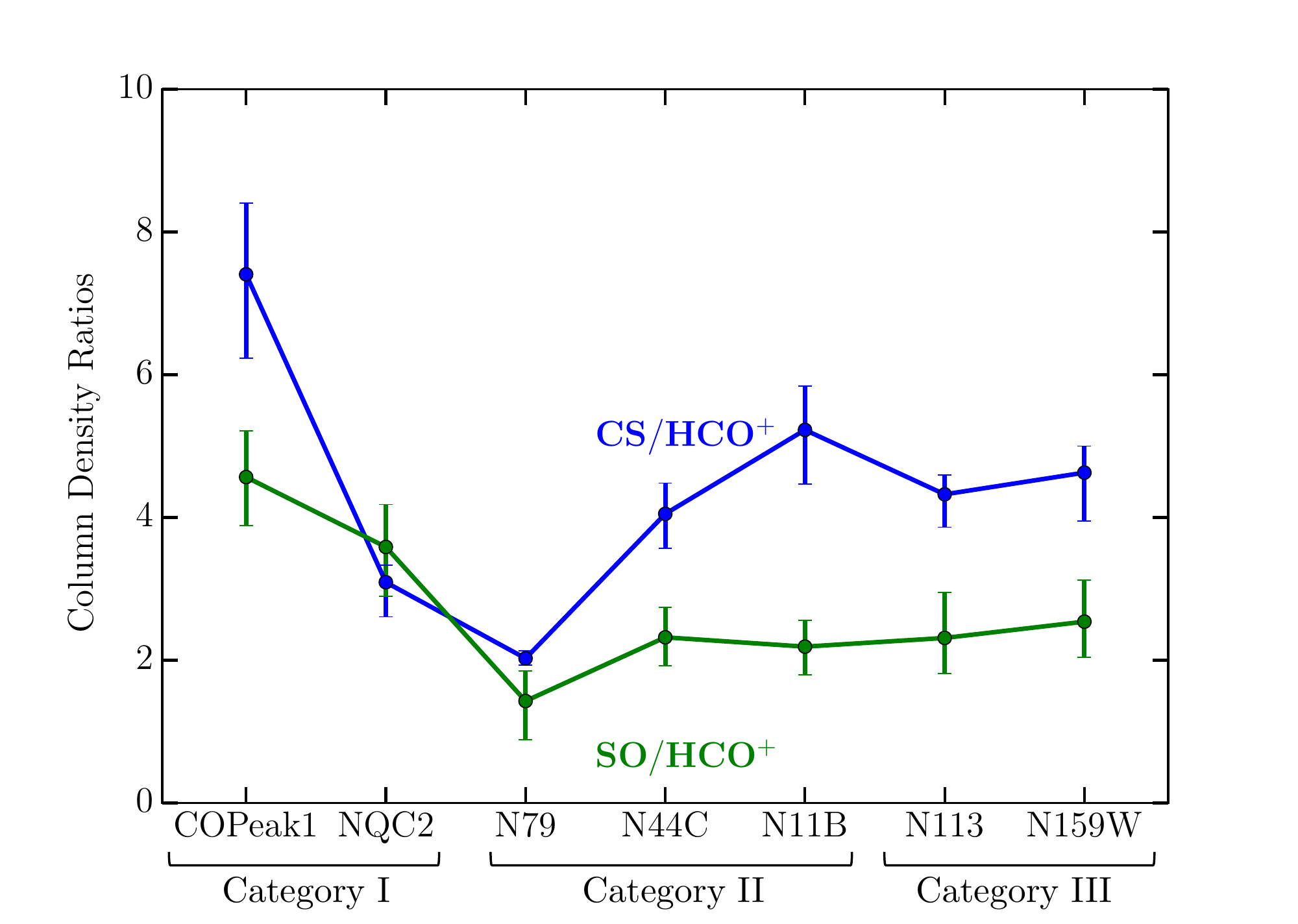}
\caption{Column density ratios of CS/HCO$^+$ and SO/HCO$^+$ in the observed sources.  
The error bars includes the errors caused by the assumed range of 
the gas kinetic temperature ($10-30$ K)
and the gas density ($5\times10^3-5\times10^4$ cm$^{-3}$).  
The CS/HCO$^+$ and SO/HCO$^+$ ratio higher in CO Peak 1 
(See Section \ref{S-bearing} for details).  
\label{s-species1}}
\end{figure}

\begin{figure}
\includegraphics[width=0.6\hsize, bb= 0 0 576 410]{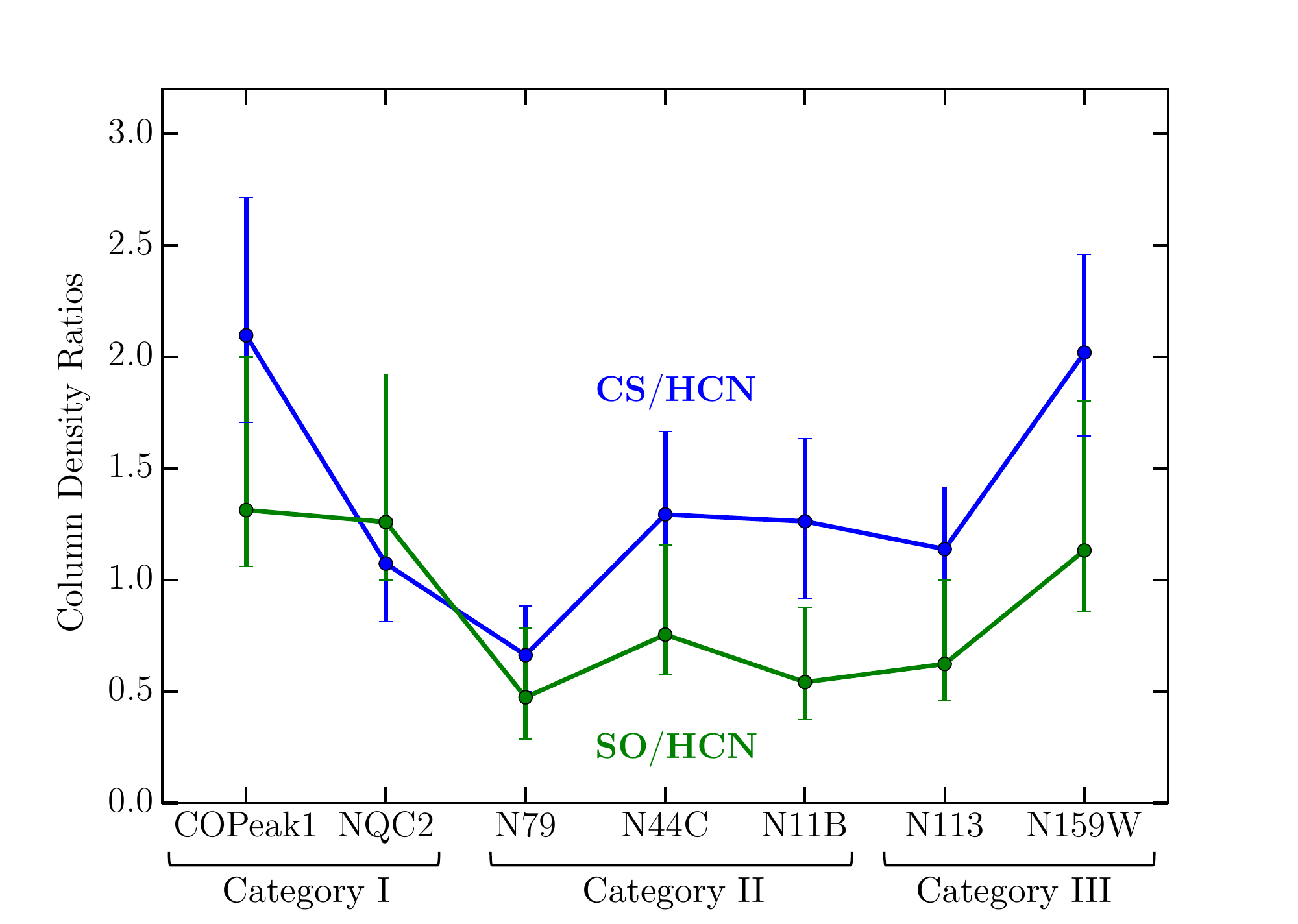}
\caption{Column density ratios of CS/HCN and SO/HCN in the observed sources.  
The error bars includes the errors caused by the assumed range of 
the gas kinetic temperature ($10-30$ K)
and the gas density ($5\times10^3-5\times10^4$ cm$^{-3}$).  
A similar trend for CO Peak 1 seen in Figure \ref{s-species1} is also seen.  
\label{s-species2}}
\end{figure}

%%%%%%%%%%%%%%%%%%%%%%%%%%%%%%%%%%%%%%%%%%%%%%%%%%
\begin{deluxetable}{lccccl}
\tabletypesize{\scriptsize}
\tablecaption{The Elemental Abundances Relative to Hydrogen Nuclei. \label{elementalabundance}}
\tablewidth{0pt}
\tablehead{
\colhead{Galaxy} & 
\colhead{${\rm O/H}\times10^4$} & 
\colhead{${\rm C/H}\times10^4$} & 
\colhead{${\rm N/H}\times10^5$} & 
\colhead{${\rm S/H}\times10^5$} & 
\colhead{References}
}
\startdata
Milky Way (Solar) & 7.41 & 4.47 &  9.12 & 1.70 & a \\
Milky Way (Solar) & 8.53 & 3.62 & 11.22 & 1.85 & a' \\
M51               & 6.31 & 3.98 & 15.85 & 1.59 & b, c \\
LMC               & 2.40 & 0.79 &  0.87 & 1.02 & a
\enddata
\tablecomments{
a) \citet{dufour1982carbon}, 
a') \citet{anders1989abundances}, 
b) \citet{bresolin2004abundance}, 
c) \citet{garnett2004CNO}.  
(The ratio of M51 is the highest one of H II regions in M51.)}
\end{deluxetable}

\clearpage

\begin{deluxetable}{lcccccccccc}
\rotate
\tabletypesize{\scriptsize}
\tablecaption{Observed Positions and Physical Properties.\label{position}}
\tablewidth{0pt}
\tablehead{
\colhead{Source Name} & \colhead{$\alpha_{\rm J2000.0}$} 
 & \colhead{$\delta_{\rm J2000.0}$} & \colhead{$V_{\rm LSR}$} 
 & \colhead{$^{12}$CO (MAGMA\tablenotemark{a})} & \colhead{$^{12}$CO (this study)}
 & \colhead{8 $\mu$m flux\tablenotemark{b}} & \colhead{24 $\mu$m flux\tablenotemark{c}}
 & \colhead{YSO} & \colhead{H II region} & \colhead{Category} \\
& \colhead{($^h$ $^m$ $^s$)} & \colhead{($^\circ$ $'$ $''$)} 
 & \colhead{(km s$^{-1}$)} & \colhead{(K km s$^{-1}$)} & \colhead{(K km s$^{-1}$)}
 & \colhead{(mJy arcsec$^{-2}$)} & \colhead{(mJy arcsec$^{-2}$)} & & & }
\startdata
CO Peak 1   \dotfill & 05 44 25.3 & $-69$ 26 25.9 & 229 & 31.1 & 31.30 &  26.8 &   11.7 & no  & no  & A \\
NQC2        \dotfill & 05 19 27.1 & $-69$ 08 48.9 & 266 & 14.4 & 13.90 &  89.0 &   58.6 & no  & no  & A \\
N79         \dotfill & 04 51 53.1 & $-69$ 23 25.3 & 232 & 17.6 & 16.80 & 946.8 & 4785.7 & yes & no  & B \\
N44C        \dotfill & 05 22 11.6 & $-67$ 58 26.0 & 282 & 21.2 & 25.58 & 387.3 & 2184.0 & yes & no  & B \\
N11B        \dotfill & 04 56 48.1 & $-66$ 24 24.2 & 285 &  9.5 & 18.74 & 332.5 & 1341.3 & yes & no  & B \\
N113        \dotfill & 05 13 18.2 & $-69$ 22 35.0 & 235 & 30.1 & 46.96 & 626.1 & 2883.8 & yes & yes & C \\
N159W       \dotfill & 05 39 36.0 & $-69$ 45 35.0 & 238 & 34.4 & 53.4  & 720.6 & 2647.5 & yes & yes & C \\
\enddata
\tablecomments{Units of right ascension are hours, minutes, and seconds, 
and units of declination are degrees, arcminutes, arcseconds.
$V_{\rm LSR}$ are of $^{13}$CO.}
\tablenotetext{a}{\citet{wong2011MAGMA}.}
\tablenotetext{b}{\textit{Spitzer} IRAC \citep{meixner2006spitzer}.}
\tablenotetext{c}{\textit{Spitzer} MIPS \citep{meixner2006spitzer}.}
\end{deluxetable}

\begin{deluxetable}{lccccll}
\rotate
\tabletypesize{\scriptsize}
\tablecaption{Achieved Sensitivity and Detected Molecules.\label{detected}}
\tablewidth{0pt}
\tablehead{
& \colhead{85-93 GHz\tablenotemark{a}} & \colhead{93-101 GHz} 
 & \colhead{100-108 GHz} & \colhead{108-116 GHz} & \\
\colhead{Source Name}  
 & \colhead{r.m.s.\tablenotemark{b}} & \colhead{r.m.s.\tablenotemark{b}}
 & \colhead{r.m.s.\tablenotemark{b}} & \colhead{r.m.s.\tablenotemark{b}} 
 & \colhead{Detected Molecules}\\
 & \colhead{(mK)} & \colhead{(mK)}
 & \colhead{(mK)} & \colhead{(mK)} & }
\startdata
CO Peak 1   \dotfill & 5.9 & 10.0 & 6.6 & 40.2 
 & CCH, HCN, HCO$^+$, HNC, \\
 &&&&& CS, SO, $^{13}$CO, $^{12}$CO \\
NQC2        \dotfill & 6.5 & 6.4 & $-$ & 20.0 
 & CCH, HCN, HCO$^+$, HNC, \\
 &&&&& CS, SO, $^{13}$CO, $^{12}$CO \\
N79         \dotfill & 13.4 & 14.4 & $-$ & 23.3
 & $c$-C$_3$H$_2$, CCH, HCN, HCO$^+$, HNC, \\
 &&&&& CS, SO, $^{13}$CO, CN, $^{12}$CO \\
N44C        \dotfill & 8.3 & 8.5 & 14.0 & 29.3
 & $c$-C$_3$H$_2$, CCH, HCN, HCO$^+$, HNC, \\
 &&&&& CS, SO, $^{13}$CO, CN, $^{12}$CO \\
N11B        \dotfill & 10.0 & 9.3 & 7.1 & 23.2  
 & $c$-C$_3$H$_2$, CCH, HCN, HCO$^+$, HNC, \\
 &&&&& CS, SO, $^{13}$CO, $^{12}$CO \\
N113        \dotfill & 22.3 & 24.3 & $-$ & 58.8 
 & $c$-C$_3$H$_2$, CCH, HCN, HCO$^+$, HNC,  \\
 &&&&& N$_2$H$^+$, CS, SO, $^{13}$CO, CN, $^{12}$CO \\
N159W       \dotfill & 14.9 & 18.9 & 25.3 & 80.9 
 & $c$-C$_3$H$_2$, CCH, HCN, HCO$^+$, HNC, \\
 &&&&& N$_2$H$^+$, CS, SO, $^{13}$CO, $^{12}$CO \\
\enddata
\tablenotetext{a}{The observed frequency is from 85 GHz to 93 GHz except for 
CO Peak 1, for which the range is from 85.25 GHz to 93.25 GHz.  }
\tablenotetext{b}{The root mean-square noise of the compressed spectra 
in the main-beam temperature ($T_{\rm MB}$) scale.}
\end{deluxetable}

\begin{deluxetable}{llrlrrrrr}
\tabletypesize{\scriptsize}
\rotate
\tablecaption{Observed Line Parameters. \label{lineparameters}}
\tablewidth{0pt}
\tablehead{
\colhead{Source Name} & \colhead{Molecule} & \colhead{Frequency} 
& \colhead{Transition} & \colhead{$T_{\rm MB}$ Peak} 
& \colhead{$V_{\rm LSR}$} & \colhead{$\Delta v$}
& \colhead{$\int T_{\rm MB}dv$} \\
\colhead{} & \colhead{} & \colhead{(GHz)} 
& \colhead{} & \colhead{(K)} & \colhead{(km s$^{-1}$)} 
& \colhead{(km s$^{-1}$)} & \colhead{(K km s$^{-1}$)}
&}
\startdata
CO Peak 1 \dotfill
& CCH & 87.284105 & $N=1-0$, $J=3/2-1/2$, $F=1-1$ &
&
&
&
$ < 0.07 $ \\
& CCH & 87.316898 & $N=1-0$, $J=3/2-1/2$, $F=2-1$ &
$ 0.019 \pm 0.006 $ &
$ 227.4 \pm 1.1 $ &
$ 6.7 \pm 2.5 $ &
$ 0.15 \pm 0.02 $ \\
& CCH & 87.328585 & $N=1-0$, $J=3/2-1/2$, $F=1-0$ &
&
&
&
$ < 0.08 $ \\
& CCH & 87.401989 & $N=1-0$, $J=1/2-1/2$, $F=1-1$ &
&
&
&
$ < 0.08 $ \\
& CCH & 87.407165 & $N=1-0$, $J=1/2-1/2$, $F=0-1$ &
&
&
&
$ < 0.08 $ \\
& CCH & 87.446470 & $N=1-0$, $J=1/2-1/2$, $F=1-0$ &
&
&
&
$ < 0.08 $ \\
& HCN & 88.631602 & $1-0$ &
$ 0.032 \pm 0.004 $ &
$ 227.2 \pm 0.7 $ &
$ 10.8 \pm 1.7 $ &
$ 0.26 \pm 0.03 $ \\
& HCO$^+$ & 89.188525 & $1-0$ &
$ 0.106 \pm 0.004 $ &
$ 226.96 \pm 0.10 $ &
$ 5.8 \pm 0.3 $ &
$ 0.652 \pm 0.012 $ \\
& HNC & 90.663568 & $1-0$ &
$ 0.040 \pm 0.005 $ &
$ 229.2 \pm 0.3 $ &
$ 5.4 \pm 0.8 $ &
$ 0.231 \pm 0.014 $ \\
& N$_2$H$^+$ & 93.173392 & $1-0$ &
&
&
&
$ < 0.04 $ \\
& CH$_3$OH & 96.741375 & $2_0-1_0$, A$^+$ &
&
&
&
$ < 0.08 $ \\
& CS & 97.980953 & $2-1$ &
$ 0.121 \pm 0.009 $ &
$ 229.28 \pm 0.18 $ &
$ 5.1 \pm 0.4 $ &
$ 0.68 \pm 0.02 $ \\
& SO & 99.299870 & $N_J=2_3-1_2$ &
$ 0.057 \pm 0.009 $ &
$ 227.6 \pm 0.4 $ &
$ 4.4 \pm 0.9 $ &
$ 0.273 \pm 0.018 $ \\
& C$^{18}$O & 109.782173 & $1-0$ &
&
&
&
$ < 0.2 $ \\
& $^{13}$CO & 110.201354 & $1-0$ &
$ 0.91 \pm 0.03 $ &
$ 230.83 \pm 0.07 $ &
$ 5.22 \pm 0.18 $ &
$ 4.95 \pm 0.07 $ \\
& CN & 113.490970 & $N=1-0$, $J=3/2-1/2$, $F=5/2-3/2$ &
&
&
&
$ < 0.4 $ \\
& $^{12}$CO & 115.271202 & $1-0$ &
$ 4.45 \pm 0.05 $ &
$ 230.51 \pm 0.03 $ &
$ 6.70 \pm 0.08 $ &
$ 31.30 \pm 0.11 $ \\
NQC2 \dotfill
& $c$-C$_3$H$_2$ & 85.338894 & $2_{12}-1_{01}$ &
&
&
&
$ < 0.06 $ \\
& CCH & 87.284105 & $N=1-0$, $J=3/2-1/2$, $F=1-1$ &
&
&
&
$ < 0.09 $ \\
& CCH & 87.316898 & $N=1-0$, $J=3/2-1/2$, $F=2-1$ &
$ 0.067 \pm 0.013 $ &
$ 266.1 \pm 0.3 $ &
$ 4.5 \pm 1.1 $ &
$ 0.343 \pm 0.020 $ \\
& CCH & 87.328585 & $N=1-0$, $J=3/2-1/2$, $F=1-0$ &
$ 0.044 \pm 0.009 $ &
$ 266.3 \pm 0.6 $ &
5 \tablenotemark{a} &
$ 0.23 \pm 0.02 $ \\
& CCH & 87.401989 & $N=1-0$, $J=1/2-1/2$, $F=1-1$ &
$ 0.047 \pm 0.009 $ &
266 \tablenotemark{a} &
$ 4.3 \pm 1.0 $ &
$ 0.217 \pm 0.017 $ \\
& CCH & 87.407165 & $N=1-0$, $J=1/2-1/2$, $F=0-1$ &
&
&
&
$ < 0.08 $ \\
& CCH & 87.446470 & $N=1-0$, $J=1/2-1/2$, $F=1-0$ &
&
&
&
$ < 0.09 $ \\
& HCN & 88.631602 & $1-0$ &
$ 0.042 \pm 0.004 $ &
$ 266.4 \pm 0.4 $ &
$ 7.5 \pm 0.8 $ &
$ 0.349 \pm 0.014 $ \\
& HCO$^+$ & 89.188525 & $1-0$ &
$ 0.141 \pm 0.004 $ &
$ 266.05 \pm 0.08 $ &
$ 4.84 \pm 0.16 $ &
$ 0.724 \pm 0.010 $ \\
& HNC & 90.663568 & $1-0$ &
$ 0.027 \pm 0.005 $ &
$ 267.7 \pm 0.6 $ &
$ 5.8 \pm 1.4 $ &
$ 0.165 \pm 0.017 $ \\
& N$_2$H$^+$ & 93.173392 & $1-0$ &
&
&
&
$ < 0.05 $ \\
& CH$_3$OH & 96.741375 & $2_0-1_0$, A$^+$ &
&
&
&
$ < 0.05 $ \\
& CS & 97.980953 & $2-1$ &
$ 0.075 \pm 0.007 $ &
$ 267.65 \pm 0.16 $ &
$ 4.0 \pm 0.5 $ &
$ 0.325 \pm 0.013 $ \\
& SO & 99.299870 & $N_J=2_3-1_2$ &
$ 0.046 \pm 0.009 $ &
$ 266.3 \pm 0.4 $ &
$ 4.9 \pm 1.1 $ &
$ 0.21 \pm 0.02 $ \\
& C$^{18}$O & 109.782173 & $1-0$ &
&
&
&
$ < 0.12 $ \\
& $^{13}$CO & 110.201354 & $1-0$ &
$ 0.675 \pm 0.014 $ &
$ 265.67 \pm 0.04 $ &
$ 3.61 \pm 0.09 $ &
$ 2.804 \pm 0.016 $ \\
& CN & 113.490970 & $N=1-0$, $J=3/2-1/2$, $F=5/2-3/2$ &
&
&
&
$ < 0.17 $ \\
& $^{12}$CO & 115.271202 & $1-0$ &
$ 2.73 \pm 0.03 $ &
$ 265.90 \pm 0.02 $ &
$ 4.68 \pm 0.05 $ &
$ 13.90 \pm 0.04 $ \\
N79 \dotfill
& $c$-C$_3$H$_2$ & 85.338894 & $2_{12}-1_{01}$ &
$ 0.046 \pm 0.008 $ &
$ 232.5 \pm 0.7 $ &
$ 7.4 \pm 1.5 $ &
$ 0.31 \pm 0.04 $ \\
& CCH & 87.284105 & $N=1-0$, $J=3/2-1/2$, $F=1-1$ &
&
&
&
$ < 0.15 $ \\
& CCH & 87.316898 & $N=1-0$, $J=3/2-1/2$, $F=2-1$ &
$ 0.152 \pm 0.016 $ &
$ 231.4 \pm 0.3 $ &
$ 5.0 \pm 0.6 $ &
$ 0.78 \pm 0.04 $ \\
& CCH & 87.328585 & $N=1-0$, $J=3/2-1/2$, $F=1-0$ &
&
&
&
$ < 0.15 $ \\
& CCH & 87.401989 & $N=1-0$, $J=1/2-1/2$, $F=1-1$ &
$ 0.072 \pm 0.014 $ &
$ 231.5 \pm 0.6 $ &
$ 6.5 \pm 1.4 $ &
$ 0.47 \pm 0.04 $ \\
& CCH & 87.407165 & $N=1-0$, $J=1/2-1/2$, $F=0-1$ &
$ 0.054 \pm 0.013 $ &
$ 231.1 \pm 0.7 $ &
5 \tablenotemark{a} &
$ 0.31 \pm 0.03 $ \\
& CCH & 87.446470 & $N=1-0$, $J=1/2-1/2$, $F=1-0$ &
&
&
&
$ < 0.14 $ \\
& HCN & 88.631602 & $1-0$ &
$ 0.155 \pm 0.007 $ &
$ 231.6 \pm 0.2 $ &
$ 9.3 \pm 0.5 $ &
$ 1.51 \pm 0.03 $ \\
& HCO$^+$ & 89.188525 & $1-0$ &
$ 0.504 \pm 0.009 $ &
$ 233.28 \pm 0.04 $ &
$ 5.22 \pm 0.11 $ &
$ 2.797 \pm 0.019 $ \\
& HNC & 90.663568 & $1-0$ &
$ 0.087 \pm 0.012 $ &
232 \tablenotemark{a} &
$ 5.1 \pm 0.8 $ &
$ 0.45 \pm 0.03 $ \\
& N$_2$H$^+$ & 93.173392 & $1-0$ &
&
&
&
$ < 0.10 $ \\
& CH$_3$OH & 96.741375 & $2_0-1_0$, A$^+$ &
&
&
&
$ < 0.10 $ \\
& CS & 97.980953 & $2-1$ &
$ 0.157 \pm 0.011 $ &
$ 233.81 \pm 0.19 $ &
$ 5.5 \pm 0.4 $ &
$ 0.95 \pm 0.03 $ \\
& SO & 99.299870 & $N_J=2_3-1_2$ &
$ 0.055 \pm 0.012 $ &
$ 231.2 \pm 0.8 $ &
$ 7.6 \pm 1.9 $ &
$ 0.46 \pm 0.05 $ \\
& C$^{18}$O & 109.782173 & $1-0$ &
&
&
&
$ < 0.17 $ \\
& $^{13}$CO & 110.201354 & $1-0$ &
$ 0.566 \pm 0.017 $ &
$ 231.03 \pm 0.07 $ &
$ 5.02 \pm 0.17 $ &
$ 3.15 \pm 0.03 $ \\
& CN & 113.490970 & $N=1-0$, $J=3/2-1/2$, $F=5/2-3/2$ &
$ 0.082 \pm 0.017 $ &
$ 233.1 \pm 0.6 $ &
5 \tablenotemark{a} &
$ 0.45 \pm 0.06 $ \\
& $^{12}$CO & 115.271202 & $1-0$ &
$ 2.90 \pm 0.04 $ &
$ 230.99 \pm 0.04 $ &
$ 5.21 \pm 0.09 $ &
$ 16.80 \pm 0.06 $ \\
N44C \dotfill
& $c$-C$_3$H$_2$ & 85.338894 & $2_{12}-1_{01}$ &
$ 0.035 \pm 0.006 $ &
$ 283.2 \pm 0.5 $ &
$ 5.6 \pm 1.1 $ &
$ 0.21 \pm 0.02 $ \\
& CCH & 87.284105 & $N=1-0$, $J=3/2-1/2$, $F=1-1$ &
&
&
&
$ < 0.7 $ \\
& CCH & 87.316898 & $N=1-0$, $J=3/2-1/2$, $F=2-1$ &
$ 0.104 \pm 0.009 $ &
$ 281.3 \pm 0.3 $ &
$ 6.9 \pm 0.7 $ &
$ 0.80 \pm 0.02 $ \\
& CCH & 87.328585 & $N=1-0$, $J=3/2-1/2$, $F=1-0$ &
$ 0.047 \pm 0.015 $ &
$ 281.8 \pm 1.0 $ &
$ 6.2 \pm 2.3 $ &
$ 0.31 \pm 0.02 $ \\
& CCH & 87.401989 & $N=1-0$, $J=1/2-1/2$, $F=1-1$ &
$ 0.053 \pm 0.009 $ &
$ 280.4 \pm 0.5 $ &
$ 5.9 \pm 1.1 $ &
$ 0.33 \pm 0.02 $ \\
& CCH & 87.407165 & $N=1-0$, $J=1/2-1/2$, $F=0-1$ &
&
&
&
$ < 0.10 $ \\
& CCH & 87.446470 & $N=1-0$, $J=1/2-1/2$, $F=1-0$ &
&
&
&
$ < 0.09 $ \\
& HCN & 88.631602 & $1-0$ &
$ 0.111 \pm 0.004 $ &
$ 281.36 \pm 0.18 $ &
$ 9.7 \pm 0.4 $ &
$ 1.192 \pm 0.017 $ \\
& HCO$^+$ & 89.188525 & $1-0$ &
$ 0.312 \pm 0.005 $ &
$ 281.29 \pm 0.06 $ &
$ 6.64 \pm 0.13 $ &
$ 2.164 \pm 0.015 $ \\
& HNC & 90.663568 & $1-0$ &
$ 0.060 \pm 0.005 $ &
$ 283.0 \pm 0.2 $ &
$ 6.1 \pm 0.6 $ &
$ 0.383 \pm 0.013 $ \\
& N$_2$H$^+$ & 93.173392 & $1-0$ &
&
&
&
$ < 0.06 $ \\
& CH$_3$OH & 96.741375 & $2_0-1_0$, A$^+$ &
&
&
&
$ < 0.06 $ \\
& CS & 97.980953 & $2-1$ &
$ 0.214 \pm 0.006 $ &
$ 282.89 \pm 0.08 $ &
$ 5.70 \pm 0.19 $ &
$ 1.324 \pm 0.015 $ \\
& SO & 99.299870 & $N_J=2_3-1_2$ &
$ 0.100 \pm 0.012 $ &
$ 281.2 \pm 0.3 $ &
$ 4.7 \pm 0.7 $ &
$ 0.51 \pm 0.03 $ \\
& C$^{18}$O & 109.782173 & $1-0$ &
&
&
&
$ < 0.11 $ \\
& $^{13}$CO & 110.201354 & $1-0$ &
$ 0.484 \pm 0.009 $ &
$ 281.41 \pm 0.05 $ &
$ 5.78 \pm 0.12 $ &
$ 2.953 \pm 0.020 $ \\
& CN & 113.490970 & $N=1-0$, $J=3/2-1/2$, $F=5/2-3/2$ &
$ 0.070 \pm 0.019 $ &
$ 281.6 \pm 0.6 $ &
$ 4.1 \pm 1.3 $ &
$ 0.34 \pm 0.03 $ \\
& $^{12}$CO & 115.271202 & $1-0$ &
$ 3.87 \pm 0.02 $ &
$ 281.533 \pm 0.016 $ &
$ 6.23 \pm 0.04 $ &
$ 25.58 \pm 0.05 $ \\
N11B \dotfill
& $c$-C$_3$H$_2$ & 85.338894 & $2_{12}-1_{01}$ &
$ 0.025 \pm 0.007 $ &
$ 287.7 \pm 0.9 $ &
$ 6.8 \pm 2.1 $ &
$ 0.19 \pm 0.03 $ \\
& CCH & 87.284105 & $N=1-0$, $J=3/2-1/2$, $F=1-1$ &
&
&
&
$ < 1.4 $ \\
& CCH & 87.316898 & $N=1-0$, $J=3/2-1/2$, $F=2-1$ &
$ 0.100 \pm 0.016 $ &
$ 283.9 \pm 0.3 $ &
$ 4.8 \pm 1.0 $ &
$ 0.51 \pm 0.03 $ \\
& CCH & 87.328585 & $N=1-0$, $J=3/2-1/2$, $F=1-0$ &
$ 0.048 \pm 0.014 $ &
$ 284.5 \pm 0.8 $ &
5 \tablenotemark{a} &
$ 0.24 \pm 0.03 $ \\
& CCH & 87.401989 & $N=1-0$, $J=1/2-1/2$, $F=1-1$ &
$ 0.056 \pm 0.012 $ &
$ 284.3 \pm 0.5 $ &
$ 5.4 \pm 1.4 $ &
$ 0.32 \pm 0.03 $ \\
& CCH & 87.407165 & $N=1-0$, $J=1/2-1/2$, $F=0-1$ &
&
&
&
$ < 0.13 $ \\
& CCH & 87.446470 & $N=1-0$, $J=1/2-1/2$, $F=1-0$ &
&
&
&
$ < 0.13 $ \\
& HCN & 88.631602 & $1-0$ &
$ 0.112 \pm 0.007 $ &
$ 284.0 \pm 0.3 $ &
$ 8.9 \pm 0.6 $ &
$ 1.10 \pm 0.02 $ \\
& HCO$^+$ & 89.188525 & $1-0$ &
$ 0.262 \pm 0.007 $ &
$ 285.80 \pm 0.07 $ &
$ 5.76 \pm 0.17 $ &
$ 1.662 \pm 0.016 $ \\
& HNC & 90.663568 & $1-0$ &
$ 0.085 \pm 0.006 $ &
$ 285.98 \pm 0.19 $ &
$ 4.2 \pm 0.3 $ &
$ 0.385 \pm 0.014 $ \\
& N$_2$H$^+$ & 93.173392 & $1-0$ &
&
&
&
$ < 0.07 $ \\
& CH$_3$OH & 96.741375 & $2_0-1_0$, A$^+$ &
&
&
&
$ < 0.07 $ \\
& CS & 97.980953 & $2-1$ &
$ 0.241 \pm 0.008 $ &
$ 285.69 \pm 0.07 $ &
$ 4.45 \pm 0.19 $ &
$ 1.151 \pm 0.014 $ \\
& SO & 99.299870 & $N_J=2_3-1_2$ &
$ 0.069 \pm 0.011 $ &
$ 284.2 \pm 0.3 $ &
$ 4.6 \pm 0.9 $ &
$ 0.36 \pm 0.02 $ \\
& C$^{18}$O & 109.782173 & $1-0$ &
&
&
&
$ < 0.16 $ \\
& $^{13}$CO & 110.201354 & $1-0$ &
$ 0.411 \pm 0.013 $ &
$ 284.65 \pm 0.08 $ &
$ 4.61 \pm 0.17 $ &
$ 2.09 \pm 0.03 $ \\
& CN & 113.490970 & $N=1-0$, $J=3/2-1/2$, $F=5/2-3/2$ &
&
&
&
$ < 0.20 $ \\
& $^{12}$CO & 115.271202 & $1-0$ &
$ 2.98 \pm 0.02 $ &
$ 284.63 \pm 0.02 $ &
$ 5.77 \pm 0.05 $ &
$ 18.74 \pm 0.04 $ \\
N113 \dotfill
& $c$-C$_3$H$_2$ & 85.338894 & $2_{12}-1_{01}$ &
$ 0.048 \pm 0.015 $ &
$ 235.3 \pm 1.3 $ &
$ 8.7 \pm 3.2 $ &
$ 0.42 \pm 0.08 $ \\
& CCH & 87.284105 & $N=1-0$, $J=3/2-1/2$, $F=1-1$ &
&
&
&
$ < 0.3 $ \\
& CCH & 87.401989 & $N=1-0$, $J=1/2-1/2$, $F=1-1$ &
$ 0.10 \pm 0.02 $ &
$ 234.6 \pm 0.8 $ &
$ 8.2 \pm 1.9 $ &
$ 0.91 \pm 0.08 $ \\
& CCH & 87.407165 & $N=1-0$, $J=1/2-1/2$, $F=0-1$ &
&
&
&
$ < 0.3 $ \\
& CCH & 87.446470 & $N=1-0$, $J=1/2-1/2$, $F=1-0$ &
&
&
&
$ < 0.2 $ \\
& HCN & 88.631602 & $1-0$ &
$ 0.328 \pm 0.015 $ &
$ 234.13 \pm 0.19 $ &
$ 8.6 \pm 0.4 $ &
$ 3.15 \pm 0.05 $ \\
& HCO$^+$ & 89.188525 & $1-0$ &
$ 0.760 \pm 0.014 $ &
$ 234.30 \pm 0.05 $ &
$ 5.41 \pm 0.12 $ &
$ 4.29 \pm 0.04 $ \\
& HNC & 90.663568 & $1-0$ &
$ 0.284 \pm 0.018 $ &
$ 236.11 \pm 0.12 $ &
$ 4.6 \pm 0.4 $ &
$ 1.39 \pm 0.03 $ \\
& N$_2$H$^+$ & 93.173392 & $1-0$ &
$ 0.056 \pm 0.009 $ &
$ 233.5 \pm 1.0 $ &
$ 13.6 \pm 2.4 $ &
$ 0.80 \pm 0.07 $ \\
& CH$_3$OH & 96.741375 & $2_0-1_0$, A$^+$ &
&
&
&
$ < 0.19 $ \\
& CS & 97.980953 & $2-1$ &
$ 0.560 \pm 0.018 $ &
$ 236.16 \pm 0.08 $ &
$ 4.65 \pm 0.17 $ &
$ 2.73 \pm 0.04 $ \\
& SO & 99.299870 & $N_J=2_3-1_2$ &
$ 0.21 \pm 0.04 $ &
$ 234.3 \pm 0.4 $ &
$ 5.2 \pm 1.1 $ &
$ 1.15 \pm 0.10 $ \\
& C$^{18}$O & 109.782173 & $1-0$ &
&
&
&
$ < 0.2 $ \\
& $^{13}$CO & 110.201354 & $1-0$ &
$ 1.49 \pm 0.02 $ &
$ 234.50 \pm 0.04 $ &
$ 4.38 \pm 0.08 $ &
$ 7.17 \pm 0.04 $ \\
& CN & 113.490970 & $N=1-0$, $J=3/2-1/2$, $F=5/2-3/2$ &
$ 0.15 \pm 0.04 $ &
$ 236.0 \pm 0.6 $ &
$ 3.9 \pm 1.4 $ &
$ 0.60 \pm 0.08 $ \\
& $^{12}$CO & 115.271202 & $1-0$ &
$ 7.95 \pm 0.04 $ &
$ 234.660 \pm 0.015 $ &
$ 5.46 \pm 0.04 $ &
$ 46.96 \pm 0.07 $ \\
N159W \dotfill
& $c$-C$_3$H$_2$ & 85.338894 & $2_{12}-1_{01}$ &
$ 0.048 \pm 0.010 $ &
$ 238.3 \pm 0.7 $ &
$ 7.4 \pm 1.8 $ &
$ 0.38 \pm 0.04 $ \\
& CCH & 87.284105 & $N=1-0$, $J=3/2-1/2$, $F=1-1$ &
&
&
&
$ < 0.19 $ \\
& CCH & 87.316898 & $N=1-0$, $J=3/2-1/2$, $F=2-1$ &
$ 0.131 \pm 0.017 $ &
$ 236.3 \pm 0.4 $ &
$ 6.7 \pm 1.0 $ &
$ 0.90 \pm 0.05 $ \\
& CCH & 87.328585 & $N=1-0$, $J=3/2-1/2$, $F=1-0$ &
$ 0.07 \pm 0.02 $ &
$ 236.3 \pm 0.9 $ &
5 \tablenotemark{a} &
$ 0.36 \pm 0.04 $ \\
& CCH & 87.401989 & $N=1-0$, $J=1/2-1/2$, $F=1-1$ &
$ 0.083 \pm 0.013 $ &
$ 235.1 \pm 0.5 $ &
5 \tablenotemark{a} &
$ 0.55 \pm 0.03 $ \\
& CCH & 87.407165 & $N=1-0$, $J=1/2-1/2$, $F=0-1$ &
&
&
&
$ < 0.19 $ \\
& CCH & 87.446470 & $N=1-0$, $J=1/2-1/2$, $F=1-0$ &
&
&
&
$ < 0.16 $ \\
& HCN & 88.631602 & $1-0$ &
$ 0.140 \pm 0.007 $ &
$ 236.1 \pm 0.3 $ &
$ 10.5 \pm 0.6 $ &
$ 1.53 \pm 0.04 $ \\
& HCO$^+$ & 89.188525 & $1-0$ &
$ 0.469 \pm 0.009 $ &
$ 237.90 \pm 0.07 $ &
$ 7.79 \pm 0.17 $ &
$ 3.87 \pm 0.03 $ \\
& HNC & 90.663568 & $1-0$ &
$ 0.153 \pm 0.011 $ &
$ 237.9 \pm 0.2 $ &
$ 6.3 \pm 0.5 $ &
$ 1.02 \pm 0.04 $ \\
& N$_2$H$^+$ & 93.173392 & $1-0$ &
$ 0.032 \pm 0.007 $ &
$ 235.7 \pm 1.2 $ &
$ 10.8 \pm 2.9 $ &
$ 0.32 \pm 0.06 $ \\
& CH$_3$OH & 96.741375 & $2_0-1_0$, A$^+$ &
&
&
&
$ < 0.13 $ \\
& CS & 97.980953 & $2-1$ &
$ 0.391 \pm 0.016 $ &
$ 237.92 \pm 0.12 $ &
$ 6.1 \pm 0.3 $ &
$ 2.54 \pm 0.04 $ \\
& SO & 99.299870 & $N_J=2_3-1_2$ &
$ 0.134 \pm 0.015 $ &
$ 236.1 \pm 0.4 $ &
$ 7.3 \pm 1.0 $ &
$ 1.02 \pm 0.06 $ \\
& C$^{18}$O & 109.782173 & $1-0$ &
&
&
&
$ < 0.9 $ \\
& $^{13}$CO & 110.201354 & $1-0$ &
$ 1.04 \pm 0.07 $ &
$ 236.6 \pm 0.2 $ &
$ 7.3 \pm 0.6 $ &
$ 7.8 \pm 0.3 $ \\
& CN & 113.490970 & $N=1-0$, $J=3/2-1/2$, $F=5/2-3/2$ &
&
&
&
$ < 1.4 $ \\
& $^{12}$CO & 115.271202 & $1-0$ &
$ 6.10 \pm 0.12 $ &
$ 236.71 \pm 0.08 $ &
$ 8.16 \pm 0.18 $ &
$ 53.4 \pm 0.4 $ \\
\enddata
\tablecomments{The errors are 1$\sigma$. 
The upper limits are 3$\sigma$.}
\tablenotetext{a}{Assumed.}
\end{deluxetable}

\clearpage

\begin{deluxetable}{lcccccccc}
\tablecaption{Correlation Coefficients of Integrated Intensities of Molecules 
among the Seven Sources. \tablenotemark{a}
\label{correlationcoefficient}}
\tablewidth{0pt}
\tablehead{
\colhead{} &
\colhead{CCH} & 
\colhead{HCN} & 
\colhead{HCO$^+$} & 
\colhead{HNC} & 
\colhead{CS} & 
\colhead{SO} & 
\colhead{$^{13}$CO} 
}
\startdata
CCH       & 1.000 & & & & & & \\
HCN       & 0.985 & 1.000 & & & & & \\
HCO$^+$   & 0.964 & 0.983 & 1.000 & & & & \\
HNC       & 0.963 & 0.966 & 0.911 & 1.000 & & & \\
CS        & 0.947 & 0.951 & 0.903 & 0.964 & 1.000 & & \\
SO        & 0.981 & 0.971 & 0.933 & 0.987 & 0.978 & 1.000 & \\
$^{13}$CO & 0.664 & 0.638 & 0.567 & 0.794 & 0.712 & 0.773 & 1.000
%CCH       & 1.000 & 0.985 & 0.964 & 0.963 & 0.947 & 0.981 & 0.664\\
%HCN       & 0.985 & 1.000 & 0.983 & 0.966 & 0.951 & 0.971 & 0.638\\
%HCO$^+$   & 0.964 & 0.983 & 1.000 & 0.911 & 0.903 & 0.933 & 0.567\\
%HNC       & 0.963 & 0.966 & 0.911 & 1.000 & 0.964 & 0.987 & 0.794\\
%CS        & 0.947 & 0.951 & 0.903 & 0.964 & 1.000 & 0.978 & 0.712\\
%SO        & 0.981 & 0.971 & 0.933 & 0.987 & 0.978 & 1.000 & 0.773\\
%$^{13}$CO & 0.664 & 0.638 & 0.567 & 0.794 & 0.712 & 0.773 & 1.000
\enddata
\tablenotetext{a}{The correlation coefficient $c$ is calculated as 
$$
c = \frac{\sum(x_i-\bar{x})(y_i-\bar{y})}{\sqrt{\sum(x_i-\bar{x})^2\sum(y_i-\bar{y})^2}} 
$$
where $x_i$ and $y_i$ are observed integrated intensities for the $i$-th source.  
$\bar{x}$ and $\bar{y}$ are the averages of $x_i$ and $y_i$, respectively.  
Correlation coefficients are higher than 0.9 except for the pairs including $^{13}$CO.  }
\end{deluxetable}

\clearpage

\begin{deluxetable}{lccccccccccc}
%\rotate
\tabletypesize{\scriptsize}
\tablecaption{Derived Column Densities\tablenotemark{a}
\label{columndensity}}
\tablewidth{0pt}
\tablecolumns{8}
\tablehead{\\[-15mm]}
\startdata
\\
& \multicolumn{11}{c}{$T=10$ K, $n_{\rm H_2}=1\times10^{4}$ cm$^{-3}$} \\
\cline{2-12}
& \multicolumn{3}{c}{Translucent Cloud} & & \multicolumn{7}{c}{LMC} \\
\cline{2-4}\cline{6-12}
& CB17 & CB24 & CB228 & & COP1 & NQC2 & N79 & N44C & N11B & N113 & N159W \\
\hline
$c$-C$_3$H$_2$ (ortho) & 84 & 3.8 & 5.4 & & - & $<4.3$ & 26 & 15 & 12 & 32 & 27 \\
CCH & 110 & 21 & 9.0 & & 31 & 73 & 190 & 170 & 120 & 400 & 210 \\
HCN & 180 & 32 & 24 & & 16 & 15 & 79 & 56 & 52 & 170 & 79 \\
HCO$^+$ & 28 & 8.8 & 2.0 & & 3.5 & 3.9 & 20 & 14 & 9.5 & 37 & 27 \\
HNC & 110 & 11 & 9.1 & & 4.1 & 3.0 & 9.0 & 7.2 & 7.3 & 34 & 22 \\
CH$_3$OH (A) & 6.2 & 2.6 & 8.0 & & $<3.3$ & $<2.1$ & $<4.2$ & $<2.5$ & $<2.9$ & $<7.9$ & $<5.4$ \\
CS & 76 & 22 & 12 & & 28 & 13 & 40 & 59 & 53 & 170 & 130 \\
SO & 34 & 33 & 38 & & 18 & 16 & 30 & 35 & 23 & 85 & 73 \\
$^{13}$CO & 7300 & 4800 & 3800 & & 4700 & 2400 & 2700 & 2700 & 1800 & 7000 & 7700 \\
\hline \\
& \multicolumn{11}{c}{$T=20$ K, $n_{\rm H_2}=1\times10^{4}$ cm$^{-3}$} \\
\cline{2-12}
& \multicolumn{3}{c}{Translucent Cloud} & & \multicolumn{7}{c}{LMC} \\
\cline{2-4}\cline{6-12}
& CB17 & CB24 & CB228 & & COP1 & NQC2 & N79 & N44C & N11B & N113 & N159W \\
\hline
$c$-C$_3$H$_2$ (ortho) & 31 & 1.3 & 1.9 & & - & $<1.8$ & 10 & 6.0 & 5.1 & 13 & 11 \\
CCH & 52 & 11 & 4.8 & & 16 & 39 & 99 & 93 & 62 & 260 & 110 \\
HCN & 81 & 15 & 11 & & 8.2 & 7.5 & 37 & 27 & 25 & 79 & 37 \\
HCO$^+$ & 13 & 4.4 & 1.1 & & 2.1 & 2.4 & 11 & 7.9 & 5.6 & 19 & 15 \\
HNC & 620 & 6.6 & 5.6 & & 2.8 & 2.0 & 6.0 & 4.8 & 4.8 & 22 & 14 \\
CH$_3$OH (A) & 3.9 & 1.6 & 8.0 & & $<2.2$ & $<1.4$ & $<2.7$ & $<1.6$ & $<1.9$ & $<5.1$ & $<3.5$ \\
CS & 38 & 11 & 6.4 & & 16 & 7.8 & 23 & 34 & 30 & 86 & 72 \\
SO & 14 & 14 & 15 & & 9.1 & 8.1 & 15 & 18 & 11 & 41 & 36 \\
$^{13}$CO & 4800 & 3600 & 3200 & & 5000 & 2600 & 3000 & 2900 & 2000 & 7100 & 8100 \\
\hline \\
& \multicolumn{11}{c}{$T=30$ K, $n_{\rm H_2}=1\times10^{4}$ cm$^{-3}$} \\
\cline{2-12}
& \multicolumn{3}{c}{Translucent Cloud} & & \multicolumn{7}{c}{LMC} \\
\cline{2-4}\cline{6-12}
& CB17 & CB24 & CB228 & & COP1 & NQC2 & N79 & N44C & N11B & N113 & N159W \\
\hline
$c$-C$_3$H$_2$ (ortho) & 18 & 0.83 & 1.1 & & - & $<1.1$ & 6.7 & 3.8 & 3.3 & 8.2 & 6.9 \\
CCH & 41 & 8.9 & 3.9 & & 13 & 31 & 80 & 75 & 50 & 210 & 92 \\
HCN & 58 & 1.1 & 7.7 & & 6.1 & 5.6 & 24 & 20 & 18 & 57 & 27 \\
HCO$^+$ & 9.9 & 3.3 & 0.91 & & 1.8 & 2.0 & 8.8 & 6.4 & 4.6 & 15 & 12 \\
HNC & 52 & 5.6 & 4.8 & & 2.5 & 1.8 & 5.3 & 4.3 & 4.2 & 18 & 12 \\
CH$_3$OH (A) & 3.4 & 1.4 & 4.3 & & $<1.9$ & $<1.2$ & $<2.4$ & $<1.4$ & $<1.7$ & $<4.5$ & $<3.1$ \\
CS & 28 & 8.2 & 4.9 & & 13 & 6.2 & 18 & 24 & 24 & 66 & 56 \\
SO & 10 & 9.9 & 11 & & 7.0 & 6.3 & 11 & 13 & 8.8 & 31 & 27 \\
$^{13}$CO & 5100 & 3900 & 2500 & & 5700 & 2900 & 3400 & 3300 & 2300 & 8100 & 9200 
\enddata
\tablenotetext{a}{In units of 10$^{12}$ cm$^{-2}$.  
The column densities are derived by the statistical equilibrium calculation (See text).  
Results are given for an H$_2$ density of $1\times10^4$ cm$^{-3}$ and three different temperatures.}
\end{deluxetable}

\clearpage

\begin{deluxetable}{lccccc}
%\rotate
\tabletypesize{\scriptsize}
\tablecaption{Column Density Ratios of Relative to HCO$^+$. 
\label{moleculeratio}}
\tablewidth{0pt}
\tablecolumns{6}
\tablehead{
 & \colhead{CCH} & \colhead{HCN} & \colhead{HNC} & \colhead{CS} & \colhead{SO}
}
\startdata
CO Peak 1   &  $8.3_{-1.3}^{+1.7}$ & $3.6_{-1.4}^{+1.3}$ & $1.2_{-0.2}^{+0.3}$ 
            &  $7.4_{-1.2}^{+1.0}$ & $4.6_{-0.7}^{+0.7}$ \\
NQC2        & $17.2_{-3.0}^{+3.7}$ & $3.0_{-1.1}^{+1.0}$ & $0.8_{-0.1}^{+0.1}$ 
            &  $3.1_{-0.5}^{+0.2}$ & $3.6_{-0.7}^{+0.6}$ \\
N79         &  $9.9_{-2.2}^{+3.1}$ & $3.2_{-0.9}^{+0.8}$ & $0.5_{-0.1}^{+0.1}$ 
            &  $2.0_{-0.1}^{+0.1}$ & $1.4_{-0.5}^{+0.4}$ \\
N44C        & $12.7_{-2.7}^{+3.3}$ & $3.2_{-1.1}^{+0.9}$ & $0.5_{-0.1}^{+0.1}$ 
            &  $4.1_{-0.5}^{+0.4}$ & $2.3_{-0.4}^{+0.4}$ \\
N11B        & $12.1_{-2.3}^{+3.1}$ & $4.3_{-1.6}^{+1.7}$ & $0.9_{-0.1}^{+0.2}$ 
            &  $5.2_{-0.8}^{+0.6}$ & $2.2_{-0.4}^{+0.4}$ \\
N113        & $13.9_{-4.4}^{+5.7}$ & $3.9_{-1.1}^{+0.9}$ & $1.0_{-0.2}^{+0.2}$ 
            &  $4.3_{-0.4}^{+0.3}$ & $2.3_{-0.5}^{+0.6}$ \\
N159W       &  $8.2_{-1.9}^{+2.6}$ & $2.3_{-0.7}^{+0.6}$ & $0.9_{-0.1}^{+0.2}$ 
            &  $4.6_{-0.8}^{+0.4}$ & $2.5_{-0.5}^{+0.6}$ \\
LMC average & $11.7\pm3.0$         & $3.4\pm0.6$         & $0.8\pm0.2$ 
            & $4.4\pm1.6$          & $2.7\pm1.0$ \\
\hline
N113\tablenotemark{a} & $-$ & 1.6 & 0.6 & 6.3 & 4.0 \\
N159W\tablenotemark{b}    & 19.3 & 0.9 & 0.3 & 1.4 & 1.6 \\
N159S\tablenotemark{b}    & 30.0 & 1.0 & 0.3 & 5.0 & 5.0 \\
N160\tablenotemark{b}     & 15.1 & 1.2 & $-$ & 1.0 & 1.7 \\
30Dor-10\tablenotemark{b} & 18.2 & 0.7 & 0.2 & 0.9 & 0.7 \\
30Dor-27\tablenotemark{b} & 38.7 & 1.2 & $-$ & 4.8 & 2.7 \\
\hline
Model of the LMC (L1)\tablenotemark{c}  & 3.4 & 2.3 & 1.3 & $-$ & $-$ \\
Model of the LMC (L2)\tablenotemark{c}  & 4.0 & 0.10 & 0.09 & $-$ & $-$ \\
Model of Milky Way (G)\tablenotemark{c} & 5.2 & 5.7 & 3.1 & $-$ & $-$ \\
\hline
Milky Way (38 translucent clouds)\tablenotemark{d} & $-$ &  9.3 & 2.2 & 8.1 & 4.5 \\
Milky Way (except CB17)\tablenotemark{d}           & $-$ & 19.3 & 3.5 & 9.5 & 10.3 \\
Milky Way (CB17, CB24, CB228)\tablenotemark{d}     & 4.2 & 6.1 & 3.4 & 5.1 & 8.4 \\
\hline
M51\tablenotemark{e} & 6.7 & 9.6 & 1.7 & 1.9 & 0.4 \\
\enddata
\tablenotetext{a}{\citet{wang2009abundances}.}
\tablenotetext{b}{\citet{heikkila1999molecular}.}
\tablenotetext{c}{\citet{millar1990chemical}.}
\tablenotetext{d}{Derived from the statistical equilibrium calculation based on 
\citet{turner1995physicsA, turner1995physicsB, turner1996physics, turner1997physics, 
turner1999physics}.}
\tablenotetext{e}{Derived from the statistical equilibrium calculation 
based on \citet{watanabe2014spectral}.}
\end{deluxetable}

\end{document}